\def\mycmd{1} 

\if\mycmd1
\documentclass[draftclsnofoot, english, a4paper, onecolumn, 12pt]{IEEEtran}
\else
\documentclass[english, tran, a4paper, twocolumn, 10pt]{IEEEtran}
\fi

\usepackage[T1]{fontenc}
\usepackage{geometry}

\if\mycmd1
\else
\fi
\usepackage{color}
\usepackage{babel}
\usepackage{verbatim}
\usepackage{float}
\usepackage{amsmath}
\usepackage{amssymb}
\usepackage{graphicx}
\usepackage[table]{xcolor}
\usepackage[utf8]{inputenc}
\usepackage{lipsum}
\usepackage{mathtools}
\usepackage{cuted}

\makeatletter

\floatstyle{ruled}
\newfloat{algorithm}{tbp}{loa}
\providecommand{\algorithmname}{Algorithm}
\floatname{algorithm}{\protect\algorithmname}

 \let\oldforeign@language\foreign@language
 \DeclareRobustCommand{\foreign@language}[1]{%
   \lowercase{\oldforeign@language{#1}}}

\usepackage{multirow}
\usepackage{amssymb}
\usepackage{array}
\usepackage{multirow}
\usepackage{graphicx}
\usepackage{amsmath}
\usepackage{latexsym}
\usepackage{stfloats}
\usepackage{color}
\usepackage{algorithm}
\usepackage{changes}
\usepackage{cite}

\newtheorem{theorem}{Theorem}
\newtheorem{lemma}{Lemma}
\newtheorem{remark}{Remark}

\newtheorem{proposition}{Proposition}

 \usepackage{subcaption}

\makeatother

\begin{document}

\title{Sum-Rate Maximization of Multicell MISO Networks with Limited Information Exchange}

\author{Youjin Kim, \IEEEmembership{Student Member, IEEE}, and Hyun Jong Yang, \IEEEmembership{Member, IEEE}\thanks{Y. Kim and H. J. Yang (corresponding author) are with the School of Electrical and Computer Engineering, Ulsan National Institute of Science and Technology (UNIST), Ulsan 44919, Republic of Korea (e-mail: \{nick0822, hjyang\}@unist.ac.kr).}}

\maketitle
\begin{abstract}
Although there have been extensive studies on transmit beamforming in multi-input single-output (MISO) multicell networks, achieving optimal sum-rate with limited channel state information (CSI) is still a challenge even with a single user per cell. A novel cooperative downlink multicell MISO beamforming scheme is proposed with highly limited information exchange among the base stations (BSs) to maximize the sum-rate.
In the proposed scheme, each BS can design its beamforming vector with only local CSI based on limited information exchange on CSI.
Unlike previous studies, the proposed beamforming design is non-iterative and does not require any \textit{vector} or \textit{matrix} feedback but requires only quantized scalar information.
The proposed scheme closely achieves the optimal sum-rate bound in almost all signal-to-noise ratio regime based on non-iterative optimization with lower amount of information exchange than existing schemes, which is justified by numerical simulations.
\end{abstract}


\IEEEpeerreviewmaketitle{}

\section{Introduction}

\label{sec:introduction}
In dense multicell networks, the signal-to-interference-plus-noise ratio (SINR) cannot grow unless the interference signals are kept weak enough compared to the desired channel gain \cite{A_Gupta15_WCL}.
If the transmitter is equipped with multiple antennas, intercell interference can be significantly mitigated or even cancelled via spatial transmit beamforming \cite{E_Bjornson10_TSP,R_Bhagavatula10_ICASSP,T_Kim16_arXiv,H_Yang17_TWC,C_Suh11_TC,E_Jorswieck08_TSP,T_Gou10_TIT,TR36.913,TR36.819,Y_Kim18_Access}.
The interference alignment framework \cite{V_Cadambe08_TIT,A_Goldsmith03_JSAC} achieves asymptotically optimal multiplexing gain based on global channel state information (CSI) at the cost of excessive use of frequency- or time-domain signal extension, but with no guarantee of optimal sum-rate achievability.
Though massive multi-input multi-output (MIMO) employed at the transmitter provide significant spectral efficiency gain \cite{L_Lu14_JSTSP,H_Tabassum16_TCom}, the number of transmit antennas even at base stations (BSs) is often limited by up to 8 in the pervasive conventional mobile networks \cite{TR38.802}.

In the downlink scenario, if information exchange for global CSI is allowed among the BSs via direct link, coordinated beamforming transmission can be employed \cite{TR36.819}.
In coordinated beamforming, only the beamforming vectors are jointly optimized, and each user's data streams are transmitted by a single serving BS.
In this paper, the focus is on the coordinated downlink multi-input single-output (MISO) beamforming design with limited direct link capacity.
With a wireless direct link, which is put on the highest priority by 3GPP, the capacity is limited by 10-100Mbps typically.
In such a case, highly limited information exchange is required, particularly in dense networks. Although the MISO multicell network is a well-studied area, achieving the optimal sum-rate with limited information exchange on CSI is still a major challenge.

\subsection{Related Works}

With global CSI, coordinated beamforming offers
optimal multiplexing gain \cite{T_Kim16_arXiv,T_Gou10_TIT,S_Jafar07_TIT},
an optimal Pareto rate boundary \cite{E_Jorswieck08_TSP}, or a significant
sum-rate gain over the conventional distributed beamforming \cite{S_He14_Elsevier,E_Larsson08_JSAC}.
However, in MISO networks, the amount of CSI information exchange in general increases as the number of transmit antennas grows, which make them difficult to be implemented in systems with limited direct link or backhaul capacity.

Several studies have proposed cooperative beamforming methods with vector quantization to reduce the amount of information exchange   \cite{F_Shi11_WCSP,H_Song10_ISCIT,J_Mirza17_TVT,O_Simeone09_EURASIP,R_Bhagavatula10_ICASSP,R_Bhagavatual11_TSP,Y_Wu14_IEICE, S_Mosleh16_TWC, S_Balaji17_Wiley, N_Lee11_TWC, Balaji18,Y_Kim18_Access,J_Kaleva17_arXiv}.
However, with the vector quantization, the number of quantization bits increases linearly with respect to the number of antennas to achieve the same rate.

Distributed beamforming also has been proposed based only on local CSI requiring no information exchange \cite{E_Bjornson10_TSP,N_Hassanpour08_ISSSTA,R_Zakhour10_TWC,E_Jorswieck08_TSP,B_Lee08_Globecom,M_Sadek07_TWC,R_Zakhour09_ITG,L_Qiang10_ICST}.
In \cite{E_Jorswieck08_TSP}, the condition of beamforming vector which corresponds to Pareto's optimal rate boundary is derived for a multicell MISO channel with local CSI. However, no closed-form solution of beamforming vector is derived. In \cite{M_Sadek07_TWC}, a simple MIMO downlink precoding is proposed in a single cell maximizing each user's signal-to-leakage-plus-noise ratio (SLNR)\footnote{The terminology is also known as signal-to-generating-interference-and-noise ratio (SGINR) \cite{B_Lee08_Globecom} or distributed virtual SINR \cite{E_Bjornson10_TSP}. } while decoupling each user's beamforming vector design. In \cite{R_Zakhour09_ITG,B_Lee08_Globecom,N_Hassanpour08_ISSSTA,L_Qiang10_ICST}, the SLNR-maximizing beamforming scheme is applied to the multicell MISO channel, and the achievability of Pareto's optimal rate bound is discussed. The same idea was extended in \cite{E_Bjornson10_TSP,R_Zakhour10_TWC} to the multicell MISO network where each user is served by all the BSs assuming each user's data being shared by all the BSs, i.e., coordinated multi-point joint transmission. Statistical beamforming design schemes robust to instantaneous CSI have also been proposed based only on the second order statistics of local CSI \cite{E_Bjornson10_TSP,H_Song10_ISCIT}.
However, the sum-rate of these SLNR-maximizing schemes with only local CSI is far below the channel capacity of the multicell MISO channel, especially in high-SNR regime.

Iterative beamforming design approaches, in which the BSs update their beamforming vectors iteratively exchanging interference pricing measures with other BSs or users, have been proposed in pursuit of maximizing the sum-rate of the two-user MIMO interference channel \cite{C_Shi09_ICC} and minimizing transmission power of the multicell MISO channel \cite{A_Tolli11_TWC,Y_Xu13_Globecom,D_Nguyen11_TSP} with the use of limited information exchange. In the scheme proposed in \cite{Q_Shi11_TSP}, beamforming vectors, receive equalizers, and weight coefficients are designed iteratively between the transmitters and receivers. However, it requires excessive amount of information exchange due to the vector information exchange about the beamforming vectors.
Furthermore, iterative optimization can significantly increase the overhead of information exchange for convergence of the solutions.

In \cite{W_Xia19_arXiv}, the beamforming vectors design based on  neural network is proposed. However, the optimal beamforming solution to the sum-rate maximization problem is still unknown.

\subsection{Contribution}

In this paper, we propose a non-iterative cooperative downlink beamforming scheme in multicell MISO networks, each cell of which consists of a BS with multiple antennas and a user with a single antenna, based on local CSI with limited information exchange of scalar values. Our contribution in summary is as follows:
\begin{itemize}
    \item We first give inspiration that the sum-rate maximization may be achieved by choosing a proper set of users and making them interference-free. From this inspiration, we propose a novel multicell beamforming design  based on the mixture of the maximization of weighted signal-to-leakage-plus-noise ratio (WSLNR) and the minimization of weighted generating-interference (WGI). Unlike previous related studies, where the SLNR or generating-interference (GI) formulation with identical weights was used, we focus on the design of the weights in WSLNR and WGI via choosing a proper set of interference-free users.
    \item For each selection on the number of interference-free users, we provide an information exchange protocol with limited direct link capacity, and present an adaptive beamforming design scheme. In the proposed protocol, only scalar information, not vector CSI, is exchanged, and hence the amount of information exchange does not grow for increasing number of antennas.
    \item Then, a scalar quantization method for the information to be exchanged is derived, based on which quantitative evaluation of the amount of information exchange is provided compared with existing schemes.
    \item We derive conditions of system parameters for which the optimal sum-rate is asymptotically achievable with the proposed scheme. We also confirm by extensive simulations that the proposed scheme closely achieves the optimal sum-rate bound for almost all the SNR regime requiring less information exchange compared to the existing schemes. Although there have been extensive studies on multicell MISO beamforming, to the best of authors' knowledge, this is the first non-iterative beamforming design that achieves the optimal sum-rate bound even with the lowest information exchange overhead.

\end{itemize}

\section{System model and Proposed Protocol}

\label{sec:System} It is assumed that each cell is composed of a single BS and user assuming frequency-, code-, or time-division multi-user orthogonal multiplexing\footnote{Though our focus is to build a beamforming design framework in case of a single user per cell, the system can be readily extended to multiuser cases, which shall be described in Section \ref{sec:multiuser}.}. Each small cell BS is assumed to
have $N_{T}$ antennas, whereas each user has a single antenna. The
number of cells considered is denoted by $N_{C}$, and it is assumed
that $N_{T}<N_{C}$ and $N_{T}\geq2$. The channel vector from the $i$-th BS (referred to as BS $i$ henceforth) to the user in the $j$-th cell (referred to as user $j$ henceforth) is denoted by $\mathbf{h}_{ij}\in\mathbb{C}^{N_{T}\times1}$.
Block fading and time-division duplexing with channel reciprocity
are assumed. Resorting to channel reciprocity, each BS is assumed
to have local CSI at the transmitter \cite{E_Bjornson10_TSP}, i.e.,
BS $i$ has the information of $\mathbf{h}_{ij}$, $j\in\left\{1,\ldots,N_{C}\right\} \triangleq\mathcal{N}_{C}$.

The beamforming vector at BS $i$ is denoted by $\mathbf{w}_{i}\in\mathbb{C}^{N_{T}\times1}$,
where $\left\Vert \mathbf{w}_{i}\right\Vert ^{2}\leq1$. The received
signal at user $i$ is written by

\begin{equation}
y_{i}=\underbrace{\mathbf{h}_{ii}^{H}\mathbf{w}_{i}x_{i}}_{\textrm{desired\,signal}}+\underbrace{\sum_{k=1,k\neq i}^{N_{C}}\mathbf{h}_{ki}^{H}\mathbf{w}_{k}x_{k}}_{\textrm{intercell\,interference}}+z_{i},
\end{equation}
where $x_{l}$ is the unit-variance transmit symbol at the $l$-th
BS, $l\in\mathcal{N}_{C}$, and $z_{i}$ is the additive white Gaussian
noise (AWGN) at user $i$ with zero-mean and variance
of $N_{0}$. Thus, the corresponding SINR is expressed by
\begin{equation}
\gamma_{i}=\left|\mathbf{h}_{ii}^{H}\mathbf{w}_{i}\right|^{2}/\left({\displaystyle \sum_{k=1,k\neq i}^{N_{C}}}\left|\mathbf{h}_{ki}^{H}\mathbf{w}_{k}\right|^{2}+N_{0}\right),\label{eq:SINR_definition}
\end{equation}
and the achievable sum-rate is given by
\begin{equation}
R=\sum_{i=1}^{N_{C}}\log(1+\gamma_{i}).
\end{equation}

\section{Optimization of the Beamforming Vector Design}
\label{sec:precoder}

\subsection{Beamforming vector design: Selection of interference-free users}
\label{sec:precoder_A}
The sum-rate maximization problem should be formulated jointly for all the beamforming vectors as
\if\mycmd1
\begin{equation}\label{eq:sum-rate max problem}
    \left( \mathbf{w}_1^{*}, \ldots, \mathbf{w}_{N_C}^{*}\right) = \arg_{\mathbf{w}_1, \ldots, \mathbf{w}_{N_C}}\displaystyle\max R\left(\mathbf{w}_1, \ldots, \mathbf{w}_{N_C}\right), \textrm{s.t.}\left\Vert \mathbf{w}_i\right\Vert ^{2}\leq1, \forall i \in \mathcal{N}_C,
\end{equation}
\else
\begin{multline}\label{eq:sum-rate max problem}
    \left( \mathbf{w}_1^{*}, \ldots, \mathbf{w}_{N_C}^{*}\right) = \arg_{\mathbf{w}_1, \ldots, \mathbf{w}_{N_C}}\displaystyle\max R\left(\mathbf{w}_1, \ldots, \mathbf{w}_{N_C}\right),
    \\ \textrm{s.t.}\left\Vert \mathbf{w}_i\right\Vert ^{2}\leq1, \forall i \in \mathcal{N}_C,
\end{multline}
\fi
which requires global CSI to find the optimal solution. According to \cite{E_Bjornson10_TSP}, the solution of the sum-rate maximization problem can also be obtained by solving the max-WSLNR problem. Specifically, let us denote the weight coefficient for the channel gain from BS $i$ to user $j$ by $\beta_{ij}\geq0$, and the set of $\beta_{ij}$, $j\in\mathcal{N}_C$, by $\boldsymbol{\beta}_i=\{\beta_{i1},\ldots,\beta_{iN_C}\}$. Then, the beamforming vector in the max-WSLNR problem for given weights is obtained from
\begin{equation}\label{eq:WSLNR_beamforming_vector_design}
    \mathbf{w}_{i, \boldsymbol{\beta}_i}  = \arg_{\mathbf{w}_i, \left\Vert \mathbf{w}_i\right\Vert ^{2}\leq1}\max \frac{\beta_{ii}\left|\mathbf{h}_{ii}^{H}\mathbf{w}_{i}\right|^2}{\sum_{j\in\mathcal{N}_{C}\setminus\{i\}}\beta_{ij}\left|\mathbf{h}_{ij}^{H}\mathbf{w}_{i}\right|^2+N_0}.
\end{equation}
Here, the weights should be jointly optimized to maximize the sum-rate as
\begin{align} \label{eq:optimal_beta}
    \left( \boldsymbol{\beta}_1^*, \ldots, \boldsymbol{\beta}_{N_C}^*\right) = \arg_{\boldsymbol{\beta}_1, \ldots, \boldsymbol{\beta}_{N_C}} \max R \left( \mathbf{w}_{1,\boldsymbol{\beta}_1}, \ldots, \mathbf{w}_{N_C,\boldsymbol{\beta}_{N_C}} \right).
\end{align}
 The problems \eqref{eq:WSLNR_beamforming_vector_design} and \eqref{eq:optimal_beta} are coupled with each other, and thus global CSI is required to solve these problems. To design the beamforming vectors with local CSI, in majority of the previous studies, all the weights are assumed to be identical, i.e., $\boldsymbol{\beta}_{i}=\boldsymbol{1}$, $\forall i\in \mathcal{N}_C$.

Our aim is to design $\boldsymbol{\beta}_{i}$, $i\in\mathcal{N}_C$, to maximize the sum-rate with local CSI and limited information exchange among the BSs. To gain intuition, we start with the following numerical example introducing the notion of \textit{interference-free users}. If the received interference at user $i$, i.e., $\sum_{k\in\mathcal{N}_C\setminus\{i\}}\left| \mathbf{h}_{ki}^{H}\mathbf{w}_k \right|^2$ in \eqref{eq:SINR_definition}, is significantly small, e.g., smaller than 1/100 of the maximum out of the interference strengths at all the users, then let us denote user $i$ by an almost-interference-free user.
Figure \ref{fig:DoF_test} shows that the optimal per-cell average rate (left y-axis) and the average number of almost-interference-free users (right y-axis) versus SNR for $N_T=4$ and $N_C=5$, where each channel is identically and independently distributed (i.i.d.) according to the complex Gaussian distribution. Here, the beamforming vectors are optimally designed through exhaustive numerical simulations based on global CSI. As shown in the figure, the average number of users with noticeably low interference increases from 0 to $N_T = 4$ as SNR increases.
The lesson from Fig. \ref{fig:DoF_test} is that choosing a proper number of interference-free users for given channel condition is essential to maximize the sum-rate. Indubitably, choosing a right set of interference-free users, i.e., who shall be interfere-free, is also critical.

\begin{center}
\begin{figure}[h]
\centering{}\includegraphics[width=\if\mycmd1 0.5 \else 0.25 \fi\paperwidth]{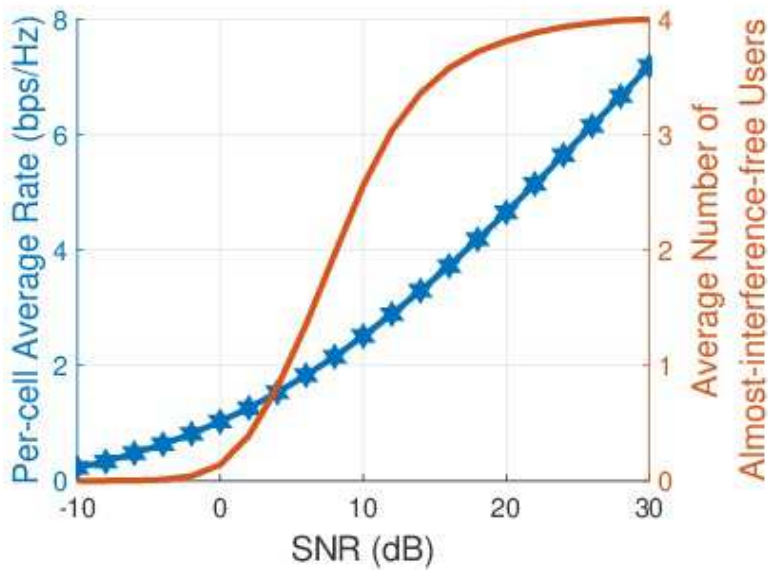}\caption{Average per-cell sum-rate and the average number of interference-free users versus SNR for $N_T=4$ and $N_C=5$}\label{fig:DoF_test}
\end{figure}
\par\end{center}

In what follows, we first propose a beamforming design framework based on the mixture of the WSLNR maximization and the WGI minimization for each possible number of interference-free users. To begin, we define a general WSLNR in pursuit of incorporating the notion of WGI as
\begin{equation}\label{eq:Xi_def}
    \chi_{i} =
    \begin{cases}
    \frac{\beta_{ii}\left|\mathbf{h}_{ii}\mathbf{w}_{i}\right|^2}{\sum_{j\in\mathcal{N}_{C}\setminus\{i\}}\beta_{ij}\left|\mathbf{h}_{ij}\mathbf{w}_{i}\right|^2+N_0} & \text{ if } \beta_{ii}\neq0\\
    \frac{1}{\sum_{j\in\mathcal{N}_{C}\setminus\{i\}}\beta_{ij}\left|\mathbf{h}_{ij}\mathbf{w}_{i}\right|^2+N_0}& \text{ if } \beta_{ii}=0,
    \end{cases}
\end{equation}
where $\beta_{ij}\geq0$ is the weight coefficient for the channel gain from BS $i$ to user $j$. The essence of the proposed beamforming design is to restrict $\beta_{ij}$ to $\beta_{ij}\in \{0,1\}$ to work with limited information exchange among the BSs. The set of interference-free users is denoted by $\mathcal{F}$, and the number of interference-free users is denoted by $\alpha$, i.e., $\left| \mathcal{F}\right| = \alpha$.

\subsection{Beamforming vector design for $\left| \mathcal{F}\right|=N_T$}
Assuming global CSI, the maximum multiplexing gain \textit{without the time or frequency domain dimension extension} can be obtained by the interference alignment framework as summarized in the following proposition.

\begin{proposition}[Theorem 1 in \cite{C_Yetis10_TSP}]\label{prop:IA_DOF}
With the interference alignment without dimension extension for the case of $N_C>N_T$, the maximum multiplexing gain is $N_T$.
\end{proposition}

Proposition \ref{prop:IA_DOF} implies that there can exist up to $N_T$ users, the effective SINRs of which after proper receive processing incorporate zero inter-user interference, i.e., $N_T$ interference-free users. Since a single antenna at the receiver is assumed, no zero-forcing-like receive processing is possible. Thus, Proposition \ref{prop:IA_DOF} in fact means that the SINRs of up to $N_T$ users can be interfere-free only via transmit beamforming.

To shed light on obtaining $N_T$ interference-free users with local CSI, we introduce the following lemma.
\begin{lemma}
\label{proposition:max_DoF}
For $N_{T}<N_{C}$, given that each BS transmits with the equal power constraint $\left\|\mathbf{w}_i\right\|^2=1$, the optimal multiplexing gain of the multicell MISO downlink channel is $N_{T}-1$ without time or frequency-domain signal extension.
\end{lemma}
\begin{IEEEproof}
Lemma \ref{proposition:max_DoF} can be proved by following the similar footsteps of \cite{C_Yetis10_TSP}. Note that the number of interference-free users is $\alpha\le N_{C}$,  and hence the multiplexing gain is $\alpha$. Suppose that user $m$ is an interference-free user. Then, the interference-free constraints at the receiver side are given by
\begin{equation}
\mathbf{h}_{km}^{H}\mathbf{w}_{k}=0,\,\,k\in\mathcal{N}_{C}\setminus\{m\}.\label{eq:interf_eq}
\end{equation}
The number of these equalities for the $\alpha$ interference-free users is $\alpha(N_{C}-1)$. On the other hand, the number of effective variables in each $\mathbf{w}_{n}$ is $N_{T}-1$ considering the unit-norm constraint. For the existence of the solution on $\mathbf{w}_{n}$ of the equalities \eqref{eq:interf_eq}, we need the number of effective variables to be equal to or greater than the number of equalities, i.e., $\alpha(N_{C}-1) \le N_{C}(N_{T}-1)\Longleftrightarrow\alpha\le\frac{N_{C}(N_{T}-1)}{N_{C}-1}$.
Therefore, the maximum number of interference-free users is given by
\begin{equation}
\alpha_{\max} =\left\lfloor \frac{N_{C}}{N_{C}-1}(N_{T}-1)\right\rfloor =N_{T}-1
\end{equation}
 for $N_{C}>N_{T}$, which proves the lemma.
\end{IEEEproof}

Lemma \ref{proposition:max_DoF} implies that the multiplexing gain of $N_T$ cannot be obtained with the equal power constraint. Inspired by this fact, we notice that $N_T$ interference-free users can be obtained by employing $\left\|\mathbf{w}_k\right\|^2=0$ for some BSs, i.e., no effective transmission. The following lemma discusses the maximum number of interference-free users based on this zero transmission power concept.
\begin{lemma} \label{lemma:max_DoF0}
The maximum number of interference-free users in the MISO interference channel with $(N_C-N_A)$ BSs having zero transmission power is given by
\begin{equation}
\alpha_{\max} =
\begin{cases}
 N_T & \text{ if } N_A=N_T \\
 N_T-1 & \text{ if } N_A>N_T \\
 N_A & \textrm{ otherwise,}
\end{cases}
\end{equation}
where $N_A$ is the number of BSs with non-zero transmission power.
\end{lemma}
\begin{IEEEproof}
 Note that the number of BSs with non-zero transmit power and the number of interference-free users having non-zero strength of the desired signal are denoted as $N_A\leq N_C$ and $\alpha\leq N_A$, respectively. The condition on $\alpha$ can be obtained following the analogous footsteps of the proof of Lemma \ref{proposition:max_DoF} by replacing $N_C$ with $N_A$ as $\alpha\le\frac{N_A(N_{T}-1)}{N_A-1}$.
Therefore, the maximum number of interference-free users is given by
\begin{align}
\alpha_{\max} & =\left\lfloor \frac{N_A}{N_A-1}(N_{T}-1)\right\rfloor =\left\lfloor \frac{(N_A-1)(N_{T}-1)+N_{T}-1}{N_A-1}\right\rfloor.
 \end{align}
 Thus, choosing $N_A=N_T$, we have $\alpha_{\max} = N_T$. Note that $\alpha_{\max}=N_T-1$ for $N_A>N_T$ and $\alpha_{\max}=N_A$ for $N_A<N_T$, which proves the lemma.
\end{IEEEproof}

From Lemma \ref{lemma:max_DoF0}, the maximum number of interference-free users, $N_T$, can be obtained by simply muting $(N_{C}-N_{T})$ BSs. In such a case, the index set of the active BSs with non-zero transmission power should be the same as the index set of the interference-free users, denoted by $\mathcal{F}$. Specifically, the beamforming vectors are designed as follows. BS $m$ for $m\in \mathcal{F}$ designs the beamforming vector that maximizes $\chi_m$ in \eqref{eq:Xi_def} setting $\beta_{mm}=0$ and $\beta_{mk}=1$ for $k\in\mathcal{F}\setminus\{m\}$, and $\beta_{mn}=0$ for $n\in\mathcal{N}_C\setminus\mathcal{F}$ as
\begin{align}
    \mathbf{w}_{m}^{\textrm{min-WGI}} & = \arg \max_{\left\|\mathbf{w}\right\|^2=1} \frac{1}{\sum_{k\in\mathcal{F}\setminus\{m\}}\left|\mathbf{h}_{mk}^{H}\mathbf{w}\right|^2+N_0}   \label{eq:w_i1} \\
    & = \arg \min_{\left\|\mathbf{w}\right\|^2=1} \left\Vert \mathbf{G}_{m}\mathbf{w}\right\Vert ^{2},
\end{align}
where $\mathbf{G}_m \triangleq \left[\sqrt{\beta_{m1}}\mathbf{h}_{m1}, \ldots, \sqrt{\beta_{mN_{C}}}\mathbf{h}_{mN_{C}}\right]^{H}$.
Then, the solution for the problem \eqref{eq:w_i1} is obtained by choosing the right singular vector of $\mathbf{G}_{m}$ associated with the smallest singular value. Note that since we choose $\beta_{mm}=0$ and $\beta_{mk}=1$ for $k\in\mathcal{F}\setminus\{m\}$, and $\beta_{mn}=0$ for $n\in\mathcal{N}_C\setminus\mathcal{F}$, the rank of  $\mathbf{G}_{m}$ is $(N_{T}-1)$; that is, the smallest singular value is 0, yielding $\left\Vert \mathbf{G}_{m}\mathbf{w}_m^{\textrm{min-WGI}}\right\Vert ^{2}=0$.

For $n\in\mathcal{N}_C\setminus\mathcal{F}$ and $\alpha = N_T$, we choose
\begin{align}
    \mathbf{w}_{n}=\boldsymbol{0}.
\end{align}
With this choice, the interference received at user $m$, $\forall m\in\mathcal{F}$, becomes zero, and the sum-rate is given by
\begin{equation}
    R=\displaystyle\sum_{m\in\mathcal{F}}\log\left(1+\frac{\left|\mathbf{h}_{mm}^{H}\mathbf{w}_{m}^{\textrm{min-WGI}}\right|^{2}}{N_{0}}\right).
\end{equation}

It is crucial to design $\mathcal{F}$ properly to maximize the sum-rate, which shall be obtained in Section \ref{sec:design set of interference-free users}.
\begin{remark}
Turning off a set of base stations in small cell networks is used as one of the sum-rate improving technologies in 3GPP \cite{TR36.872}. However, which and how many BSs should be turned off to maximize the sum-rate for given network has been investigated only empirically or heuristically. In this study, we derive which and how many BSs should be turned off in case of $\left|\mathcal{F}\right|=N_T$ to nearly achieve the maximum capacity bound.
\end{remark}

\subsection{Beamforming vector design for $\left| \mathcal{F}\right| =N_T-1$}
From Lemma \ref{lemma:max_DoF0}, $\left| \mathcal{F}\right| =\alpha=N_T-1$ can be obtained by having $N_A\ge N_T-1$. Setting $N_A$ to its maximum value, i.e., $N_A=N_C$, does not harm the sum-rate because more non-zero rates from BS $n$, $n\in\mathcal{N}_C\setminus\mathcal{F}$, are added in the sum-rate than with $N_A<N_C$. Thus, for $\alpha=N_T-1$, we choose to set $N_A=N_C$. For $\alpha=N_T-1$, we consider the following beamforming designs with local CSI.

\subsubsection{BS $n$ for $n\in\mathcal{N}_C\setminus\mathcal{F}$}
Note that each beamforming vector of size $N_T$ has null space size of $N_T-1$. Thus, to make user $m$, $m\in\mathcal{F}$, interference-free, BS $n$, $n\in\mathcal{N}_C\setminus\mathcal{F}$ should employ the min-WGI beamforming design in \eqref{eq:w_i1} as follows:
\begin{align} \label{eq:minWGI_w1}
    \mathbf{w}_{n}^{\textrm{min-WGI}} = \arg \min_{\left\|\mathbf{w}\right\|^2=1} \sum_{m\in\mathcal{F}}\left|\mathbf{h}_{nm}^{H}\mathbf{w}\right|^2.
\end{align}

\subsubsection{BS $m$ for $m\in\mathcal{F}$}
Since BS $m$ for $m\in\mathcal{F}$ only needs to make zero interference to the BSs with the indices in $\mathcal{F}\setminus\{m\}$, where $\left|\mathcal{F}\setminus\{m\}\right|=N_T-2$, BS $m$ can utilize the space of rank one either to improve the desired channel gain or to make zero-interference to user $l$ for $l\in\mathcal{N}_C\setminus\mathcal{F}$.
Specifically, to make zero-interference to user $l$ for $l\in\mathcal{N}_C\setminus\mathcal{F}$, BS $m$ for $m\in\mathcal{F}$ would set $\beta_{mq}=1$ for $q\in\left(\mathcal{F}\cup \{l\}\right)\setminus\{m\}$ and $\beta_{mm}=\beta_{mn}=0$ for $n\in\mathcal{N}_C\setminus\mathcal{F}$ and design its beamforming vector maximizing \eqref{eq:Xi_def} from
\begin{align} \label{eq:minWGI_w}
    \mathbf{w}_{m}^{\textrm{min-WGI}} = \arg \min_{\left\|\mathbf{w}\right\|^2=1} \sum_{q\in\left(\mathcal{F}\cup \{l\}\right)\setminus\{ m\}}\left|\mathbf{h}_{mq}^{H}\mathbf{w}\right|^2.
\end{align}
On the other hand, to improve the desired channel gain, BS $m$ for $m\in\mathcal{F}$ would set $\beta_{mm}=\beta_{mk}=1$ for $k\in\mathcal{F}\setminus\{m\}$ and $\beta_{mn}=0$ for $n\in\mathcal{N}_C\setminus\mathcal{F}$ and design its beamforming vector maximizing \eqref{eq:Xi_def} as
\begin{align}  \label{eq:maxWSLNR_w0}
    \mathbf{w}_{m}^{\textrm{max-WSLNR}} &= \arg \max_{\left\|\mathbf{w}\right\|^2=1} \frac{\left|\mathbf{h}_{mm}^{H}\mathbf{w}\right|^2}{\sum_{k\in\mathcal{F}\setminus\{m\}}\left|\mathbf{h}_{mk}^{H}\mathbf{w}\right|^2+N_0}\\
    &=  \arg \max_{\left\|\mathbf{w}\right\|^2=1} \frac{\mathbf{w}^{H}\mathbf{A}_{m}\mathbf{w}}{\mathbf{w}^{H}\mathbf{B}_{m}\mathbf{w}},
\end{align}
where $\mathbf{A}_{m}=\mathbf{h}_{mm}\mathbf{h}_{mm}^{H}$ and $\mathbf{B}_{m}={\displaystyle \sum_{k\in\mathcal{F}\setminus\{m\}}}\mathbf{h}_{mk}\mathbf{h}_{mk}^{H}+N_{0}\mathbf{I}$.
Then, the solution of \eqref{eq:maxWSLNR_w0} is given by the eigenvector of $\mathbf{B}_{m}^{-1}\mathbf{A}_{m}$ associated with the maximum eigenvalue.

To discuss the difference between the aforementioned two strategies in the sense of maximizing the sum-rate, we establish the following theorem.
\begin{theorem} \label{theorem:R1_R2}
For BS $m$, $m\in\mathcal{F}$, and $\alpha= N_T-1$, let us denote the sum-rate for the case (referred to as Case 1) where BS $m$ employs the max-WSLNR beamforming from \eqref{eq:maxWSLNR_w0} as $R_{1}$, and the sum-rate for the case (referred to as Case 2) where BS $m$ employs the min-WGI beamforming from \eqref{eq:minWGI_w} by $R_{2}$. For both the cases, BS $n$, $n\in\mathcal{N}_{C}\setminus\mathcal{F}$, designs its beamforming vector from \eqref{eq:minWGI_w1}. Then, we have $R_{1}-R_{2}\geq0$ in low- and high-SNR regime.
\end{theorem}
\begin{IEEEproof}
The sum-rate for Case 1, $R_{1}$, can be represented as
\if\mycmd1
\begin{equation}
    R_{1} =\sum_{m\in\mathcal{F}}\log\left(1+\frac{\tilde{\eta}_{mm}^{[1]}}{N_{0}}\right) + \sum_{n\in\mathcal{N}_{C}\setminus\mathcal{F}}\log\left(1+\frac{\tilde{\eta}_{nn}^{[2]}}{{\displaystyle \sum_{m\in\mathcal{F}}}\tilde{\eta}_{mn}^{[1]}+{\displaystyle \sum_{v\in\mathcal{N}_{C}\setminus\mathcal{F}, v\neq n}}\tilde{\eta}_{vn}^{[2]}+N_{0}}\right),
\end{equation}
\else
\begin{multline}
R_{1} =\sum_{m\in\mathcal{F}}\log\left(1+\frac{\tilde{\eta}_{mm}^{[1]}}{N_{0}}\right)
\\ + \sum_{n\in\mathcal{N}_{C}\setminus\mathcal{F}}\log\left(1+\frac{\tilde{\eta}_{nn}^{[2]}}{{\displaystyle \sum_{m\in\mathcal{F}}}\tilde{\eta}_{mn}^{[1]}+{\displaystyle \sum_{v\in\mathcal{N}_{C}\setminus\mathcal{F}, v\neq n}}\tilde{\eta}_{vn}^{[2]}+N_{0}}\right),
\end{multline}
\fi
where $\tilde{\eta}_{ij}^{[1]}=\left|\mathbf{h}_{ij}^{H}\mathbf{w}_{i}^{\textrm{max-WSLNR}}\right|^{2}$ and $\tilde{\eta}_{ij}^{[2]}=\left|\mathbf{h}_{ij}^{H}\mathbf{w}_{i}^{\textrm{min-WGI}}\right|^{2}$.

To compute $R_{2}$, suppose that BSs $m$, $m\in\mathcal{F}$, make GI to another user $l$, $l\in\mathcal{N}_{C}\setminus\mathcal{F}$, zero. Then, $R_{2}$ can be represented as
\if\mycmd1
\begin{equation}
    R_{2} =\sum_{m\in\mathcal{F}}\log\left(1+\frac{\tilde{\eta}_{mm}^{[2]}}{N_{0}}\right) + \log\left(1+\frac{\tilde{\eta}_{ll}^{[2]}}{{\displaystyle \sum_{g\in\mathcal{N}_{C}\setminus\mathcal{F},g\neq l}}\tilde{\eta}_{gl}^{[2]}+N_{0}}\right) + \sum_{n\in\mathcal{N}_{C}\setminus\mathcal{F},n\neq l}\log\left(1+\frac{\tilde{\eta}_{nn}^{[2]}}{{\displaystyle \sum_{h\in\mathcal{N}_{C}\setminus\{n\}}}\tilde{\eta}_{hn}^{[2]}+N_{0}}\right).
\end{equation}
\else
\begin{multline}
R_{2} =\sum_{m\in\mathcal{F}}\log\left(1+\frac{\tilde{\eta}_{mm}^{[2]}}{N_{0}}\right) + \log\left(1+\frac{\tilde{\eta}_{ll}^{[2]}}{{\displaystyle \sum_{g\in\mathcal{N}_{C}\setminus\mathcal{F},g\neq l}}\tilde{\eta}_{gl}^{[2]}+N_{0}}\right)\\
 + \sum_{n\in\mathcal{N}_{C}\setminus\mathcal{F},n\neq l}\log\left(1+\frac{\tilde{\eta}_{nn}^{[2]}}{{\displaystyle \sum_{h\in\mathcal{N}_{C}\setminus\{n\}}}\tilde{\eta}_{hn}^{[2]}+N_{0}}\right).
\end{multline}
\fi

i) In low-SNR regime, i.e., $N_{0}$ is arbitrarily large,
\begin{equation}
R_{1}\simeq\sum_{m\in\mathcal{F}}\log\left(1+\frac{\tilde{\eta}_{mm}^{[1]}}{N_{0}}\right)+\sum_{n\in\mathcal{N}_{C}\setminus\mathcal{F}_{c}}\log\left(1+\frac{\tilde{\eta}_{nn}^{[2]}}{N_{0}}\right),
\end{equation}
\begin{equation}
R_{2}\simeq\sum_{m\in\mathcal{F}}\log\left(1+\frac{\tilde{\eta}_{mm}^{[2]}}{N_{0}}\right)+\sum_{n\in\mathcal{N}_{C}\setminus\mathcal{F}}\log\left(1+\frac{\tilde{\eta}_{nn}^{[2]}}{N_{0}}\right).
\end{equation}
Consequently, we have
\begin{align}
R_{1}-R_{2} & \simeq\sum_{m\in\mathcal{F}}\log\left(1+\frac{\tilde{\eta}_{mm}^{[1]}}{N_{0}}\right)-\sum_{m\in\mathcal{F}}\log\left(1+\frac{\tilde{\eta}_{mm}^{[2]}}{N_{0}}\right),\label{eq:R_WSLRN-R_WGI}\\
 & \simeq\sum_{m\in\mathcal{F}}\log\left(1+\frac{\mbox{\ensuremath{\left\Vert \mathbf{h}_{mm}\right\Vert }}^{2}}{N_{0}}\right)-\sum_{m\in\mathcal{F}}\log\left(1+\frac{\tilde{\eta}_{mm}^{[2]}}{N_{0}}\right),\label{eq:R_diff_2}
\end{align}
where \eqref{eq:R_diff_2} follows from the fact that the max-WSLNR problem \eqref{eq:maxWSLNR_w0} becomes the max-SNR problem for arbitrarily large $N_{0}$, yielding $\tilde{\eta}_{mm}^{[1]}=\left|\mathbf{h}_{mm}^{H}\mathbf{w}_{m}^{\textrm{max-WSLNR}}\right|^{2}\simeq\mbox{\ensuremath{\left\Vert \mathbf{h}_{mm}\right\Vert }}^{2}$.
Since $\mbox{\ensuremath{\left\Vert \mathbf{h}_{mm}\right\Vert }}^{2}\ge\left|\mathbf{h}_{mm}^{H}\mathbf{w}\right|^{2}$ for any unit-norm $\mathbf{w}$, we have $\mbox{\ensuremath{\left\Vert \mathbf{h}_{mm}\right\Vert }}^{2}\ge\tilde{\eta}_{mm}^{[2]}$ for $m\in\mathcal{F}$, which proves the theorem for low-SNR regime.

ii) In high-SNR regime, i.e., $N_{0}$ is arbitrarily small, the achievable rates of the interference-free users, which have zero interference, are dominant due to the interference terms in the achievable rates of the other users. Thus, we have $R_{1}\simeq\sum_{m\in\mathcal{F}_{c}}\log\left(1+\frac{\tilde{\eta}_{mm}^{[1]}}{N_{0}}\right)$ and $R_{2}\simeq\sum_{m\in\mathcal{F}_{c}}\log\left(1+\frac{\tilde{\eta}_{mm}^{[2]}}{N_{0}}\right)$, and hence, we again have the same $R_{1}-R_{2}$ expression as in \eqref{eq:R_WSLRN-R_WGI}. In Case 1, $\mathbf{w}_{m}$ for $m\in\mathcal{F}$ is designed to have the direction of the orthogonal projection of $\mathbf{h}_{mm}$ onto the null space of $\mathbf{h}_{mn}$, $n\in\mathcal{F}\setminus\{m\}$. On the other hand, in Case 2, the beamforming vector is designed to have the direction of the null space of $\mathbf{h}_{mn}$ and $\mathbf{h}_{ml}$. That is, the beamforming vector is designed independently of $\mathbf{h}_{mm}$ on the null space of $\mathbf{h}_{mn}$, $n\in\mathcal{F}\setminus\{m\}$.
Therefore, we have
\begin{equation}
\tilde{\eta}_{mm}^{[1]}=\left|\mathbf{h}_{mm}^{H}\mathbf{w}_{m}^{\textrm{max-WSLNR}}\right|^{2}\geq\tilde{\eta}_{mm}^{[2]}=\left|\mathbf{h}_{mm}^{H}\mathbf{w}_{m}^{\textrm{min-WGI}}\right|^{2},
\end{equation}
which proves the theorem for high-SNR regime.
\end{IEEEproof}

From Theorem \ref{theorem:R1_R2}, we propose to design $\mathbf{w}_m$ for $\alpha=N_T-1$, $m\in\mathcal{F}$, from the max-WSLNR problem of \eqref{eq:maxWSLNR_w0}.The sum-rate with such a choice is given by \eqref{eq:sumrate_N_T-1}.

\if\mycmd1
\begin{multline}\label{eq:sumrate_N_T-1}
R = \underbrace{\displaystyle\sum_{m\in\mathcal{F}}\log\left(1+\frac{\left|\mathbf{h}_{mm}^{H}\mathbf{w}_{m}^{\textrm{max-WSLNR}}\right|^{2}}{N_{0}}\right)}_{\text{no received interference}} \\ + \underbrace{\displaystyle\sum_{n\in\mathcal{N}_C\setminus\mathcal{F}}\log\left(1+\frac{\left|\mathbf{h}_{nn}^{H}\mathbf{w}_{n}^{\textrm{min-WGI}}\right|^{2}}{\displaystyle\sum_{m\in\mathcal{F}}\left|\mathbf{h}_{mn}^{H}\mathbf{w}_{m}^{\textrm{max-WSLNR}}\right|^{2} + \displaystyle\sum_{v\in\mathcal{N}_C\setminus\mathcal{F},v\neq n}\left|\mathbf{h}_{vn}^{H}\mathbf{w}_{v}^{\textrm{min-WGI}}\right|^{2} + N_{0}}\right)}_{\text{includes interference received from all the BSs}}
\end{multline}
\else
\begin{figure*}
\begin{equation}\label{eq:sumrate_N_T-1}
R = \underbrace{\displaystyle\sum_{m\in\mathcal{F}}\log\left(1+\frac{\left|\mathbf{h}_{mm}^{H}\mathbf{w}_{m}^{\textrm{max-WSLNR}}\right|^{2}}{N_{0}}\right)}_{\text{no received interference}} + \underbrace{\displaystyle\sum_{n\in\mathcal{N}_C\setminus\mathcal{F}}\log\left(1+\frac{\left|\mathbf{h}_{nn}^{H}\mathbf{w}_{n}^{\textrm{min-WGI}}\right|^{2}}{\displaystyle\sum_{m\in\mathcal{F}}\left|\mathbf{h}_{mn}^{H}\mathbf{w}_{m}^{\textrm{max-WSLNR}}\right|^{2} + \displaystyle\sum_{v\in\mathcal{N}_C\setminus\mathcal{F},v\neq n}\left|\mathbf{h}_{vn}^{H}\mathbf{w}_{v}^{\textrm{min-WGI}}\right|^{2} + N_{0}}\right)}_{\text{includes interference received from all the BSs}}
\end{equation}
\end{figure*}
\fi
Again, the design of $\mathcal{F}$ shall be provided in Section \ref{sec:design set of interference-free users}.

\subsection{Beamforming vector design for $\left|\mathcal{F} \right|\leq N_T-2$}
For $\left|\mathcal{F} \right|=\alpha\leq N_T-2$, all the BSs design their beamforming vectors making zero GI to user $m$, $m\in\mathcal{F}$. The number of neighboring users, to which each BS makes GI zero, is $\alpha-1$ for BS $m$, $m\in\mathcal{F}$, and $\alpha$ for BS $n$, $n\in\mathcal{N}_C\setminus\mathcal{F}$. That is, BS $m$, $m\in\mathcal{F}$, designs its beamforming vector maximizing the desired channel gain and making GI zero to user $k$, $k\in\mathcal{F}\setminus\{m\}$, and BS $n$, $n\in \mathcal{N}_C\setminus\mathcal{F}$, designs its beamforming vectors maximizing the desired channel gain and making GI zero to user $m$, $m\in\mathcal{F}$. Then, the beamforming vectors of BS $m$ and the $n$-BS are designed in the null spaces of ranks $(N_T-\alpha+1)$ and $(N_T-\alpha)$, respectively. Then, for $\alpha\leq N_T-2$, all the beamforming vectors are obtained from the max-WSLNR problem of \eqref{eq:maxWSLNR_w0}. The sum-rate in such a case is given by
\if\mycmd1
\begin{equation}
    R = \underbrace{\displaystyle\sum_{m\in\mathcal{F}}\log\left(1+\frac{\left|\mathbf{h}_{mm}^{H}\mathbf{w}_{m}^{\textrm{max-WSLNR}}\right|^{2}}{N_{0}}\right)}_{\text{no received interference}} + \underbrace{\displaystyle\sum_{n\in\mathcal{N}_C\setminus\mathcal{F}}\log\left(1+\frac{\left|\mathbf{h}_{nn}^{H}\mathbf{w}_{n}^{\textrm{max-WSLNR}}\right|^{2}}{\displaystyle\sum_{h\in\mathcal{N}_C\setminus\{n\}}\left|\mathbf{h}_{hn}^{H}\mathbf{w}_{h}^{\textrm{max-WSLNR}}\right|^{2} + N_{0}}\right)}_{\text{includes interference received from all the BSs}}.
\end{equation}
\else
\begin{multline}
    R = \underbrace{\displaystyle\sum_{m\in\mathcal{F}}\log\left(1+\frac{\left|\mathbf{h}_{mm}^{H}\mathbf{w}_{m}^{\textrm{max-WSLNR}}\right|^{2}}{N_{0}}\right)}_{\text{no received interference}} + \\ \underbrace{\displaystyle\sum_{n\in\mathcal{N}_C\setminus\mathcal{F}}\log\left(1+\frac{\left|\mathbf{h}_{nn}^{H}\mathbf{w}_{n}^{\textrm{max-WSLNR}}\right|^{2}}{\displaystyle\sum_{h\in\mathcal{N}_C\setminus\{n\}}\left|\mathbf{h}_{hn}^{H}\mathbf{w}_{h}^{\textrm{max-WSLNR}}\right|^{2} + N_{0}}\right)}_{\text{includes interference received from all the BSs}}.
\end{multline}
\fi

\begin{figure}
    \centering
    \begin{subfigure}[b]{\if\mycmd1 0.5 \else 0.25 \fi\paperwidth}
        \includegraphics[width=\textwidth]{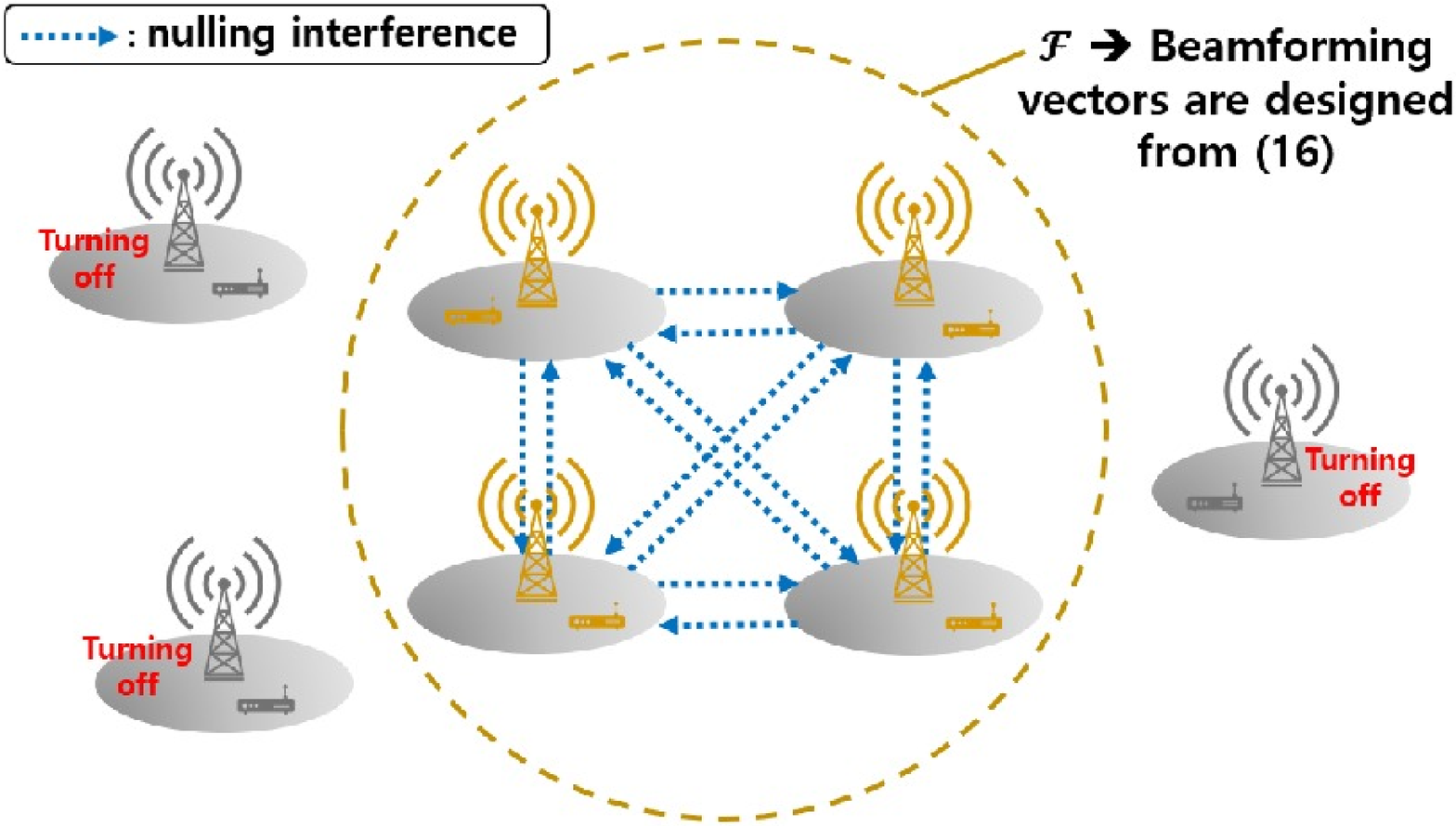}
        \caption{$\alpha=N_T$}
        \label{fig:beamforming Nt}
    \end{subfigure}
        \begin{subfigure}[b]{\if\mycmd1 0.5 \else 0.25 \fi\paperwidth}
        \includegraphics[width=\textwidth]{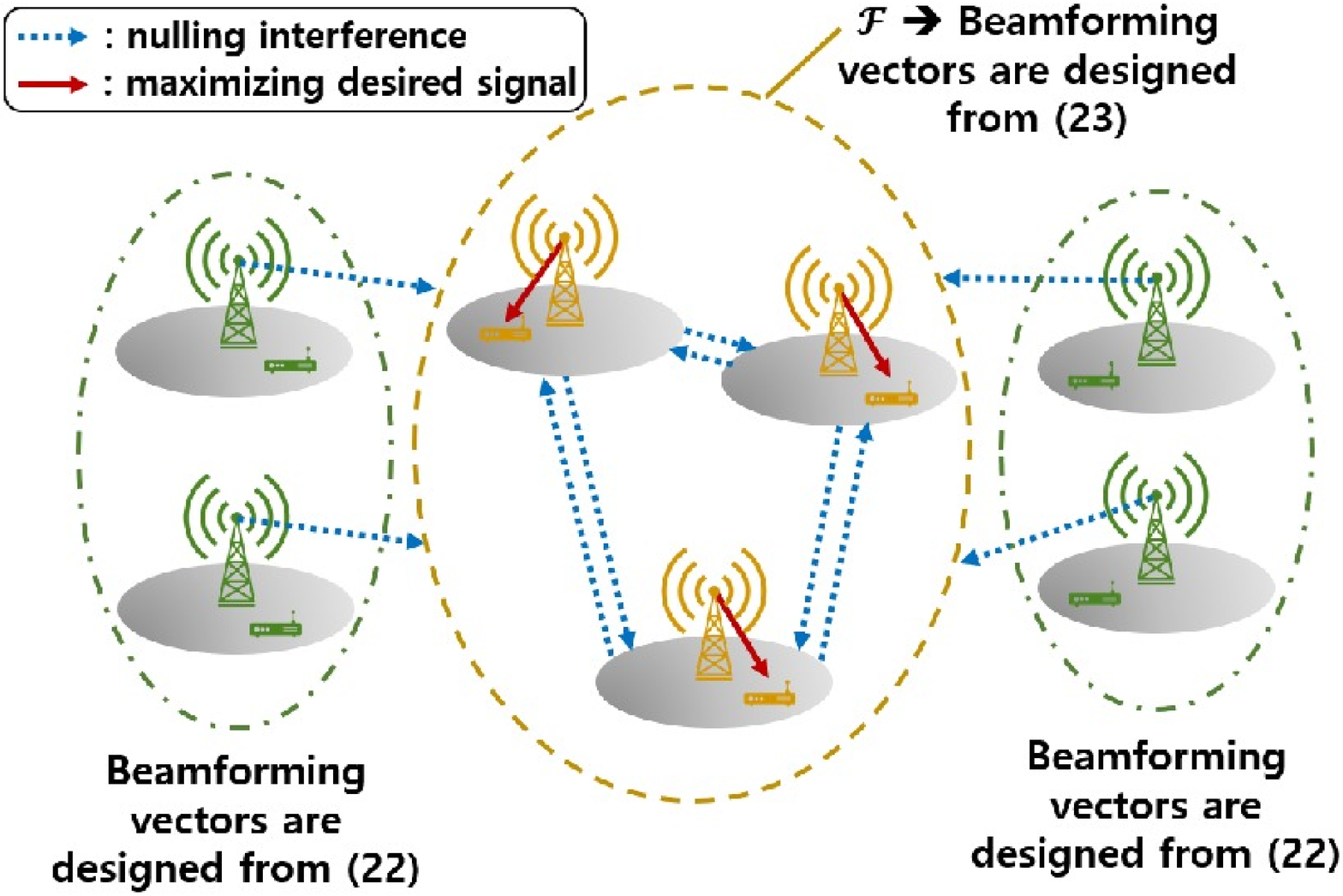}
        \caption{$\alpha = N_T-1$}
        \label{fig:beamforming Nt-1}
    \end{subfigure}
        \begin{subfigure}[b]{\if\mycmd1 0.5 \else 0.25 \fi\paperwidth}
        \includegraphics[width=\textwidth]{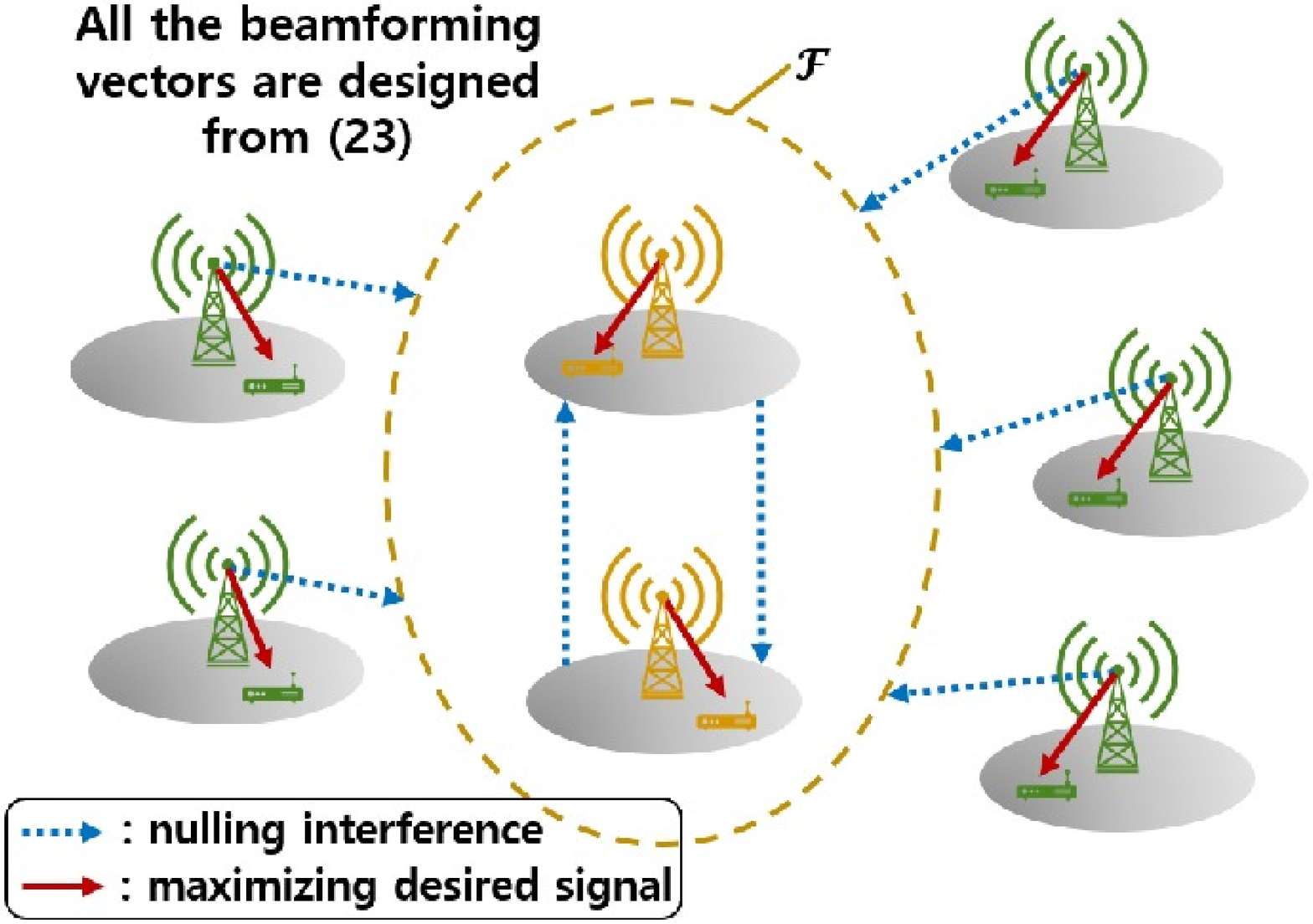}
        \caption{$\alpha = N_T-2$}
        \label{fig:beamforming Nt-2}
    \end{subfigure}
        \caption{Illustration of the proposed multicell beamforming vector design}
\end{figure}

The examples of the beamforming vector design protocol with $N_T=4$ and $N_C=7$ for $\alpha=N_T=4$, $\alpha=N_T-1=3$, and $\alpha=N_T-2=2$ are illustrated in Figs. \ref{fig:beamforming Nt},  \ref{fig:beamforming Nt-1}, and \ref{fig:beamforming Nt-2}, respectively.

\subsection{Selection of $\mathcal{F}$: Design of $\beta_{ij}$} \label{sec:design set of interference-free users}
Now, the aim is to determine a proper number of interference-free users, $\alpha$, and the set of interference-free users, $\mathcal{F}$, out of all possible cases in pursuit of maximizing the sum-rate with local CSI and limited information exchange. Totally, there exist $N_{K} = {N_C \choose N_T} + {N_C \choose N_T-1} + \cdots + {N_C \choose 1}$ possible interference-free users selection.

Let us denote the $c$-th interference-free user selection, $c\in\{1,\ldots,N_{K}\}$, by $\mathcal{F}_{c}$, i.e., users $m$, $m\in\mathcal{F}_{c}$, have received interference of zero.
With this interference-free users selection of $\mathcal{F}_{c}$, the rate of user $i$ is denoted by $r_i^{[c]}$ and the beamforming vector of BS $i$ is denoted by $\mathbf{w}_{i}^{[c]}$.
Then, the sum-rate for the $c$-th interference-free user selection can be represented as
\begin{equation}
R^{[c]} =\underbrace{\sum_{m\in\mathcal{F}_{c}}\log\left(1+\frac{\eta_{mm}^{[c]}}{N_{0}}\right)}_{\triangleq R_{\textrm{local}}^{[c]}} +\underbrace{\sum_{n\in\mathcal{N}_{C}\setminus\mathcal{F}_{c}}\log\left(1+\frac{\eta_{nn}^{[c]}}{T_n^{[c]}}\right)}_{\triangleq R_{\textrm{global}}^{[c]}}, \label{eq:R_c}
\end{equation}
where $\eta_{ij}^{[c]}=\left|\mathbf{h}_{ij}^{H}\mathbf{w}_{i}^{[c]}\right|^{2}$ and $T_n^{[c]} = {\displaystyle \sum_{k\in\mathcal{N}_{C}\setminus\{n\}}}\eta_{kn}^{[c]}+N_{0}$.
Herein, the first term $R_{\textrm{local}}^{[c]}$ is the sum of rates of user $m$, $m\in\mathcal{F}_c$, which can be computed with only local CSI by BS $m$.
On the other hand, the second term $R_{\textrm{global}}^{[c]}$ is the sum of rates of user $n$ which require global CSI to be computed by BS $n$, $n\in\mathcal{N}_C\setminus\mathcal{F}_c$.

For $\alpha=N_T$, $R_{\textrm{global}}^{[c]}$ is zero since BS $n$, $n\in\mathcal{N}_C\setminus\mathcal{F}$, has zero transmit power.
Therefore, we have $R^{[c]}=R_{\textrm{local}}^{[c]}$ and it requires only local CSI to be available at BS $m$, $m\in\mathcal{F}_c$. However, for $\alpha\leq N_T-1$, $R_{\textrm{global}}^{[c]}$ is non-zero and requires global CSI to be available at all the BSs. Thus, we propose to consider the upper bound of the average $R_{\textrm{global}}^{[c]}$ which can be computed at all the BSs with only local CSI.
To get the upper bound of $E\{ R_{\textrm{global}}^{[c]}\}$ for $\alpha\leq N_T-1$, we establish the following lemma.
\begin{lemma} \label{lemma:R_global_upper}
For all $c\in\{1,\ldots,N_K\}$ and $\alpha\leq N_T-1$,
\if\mycmd1
\begin{equation}\label{eq:R_global_upper}
    \mathrm{E}\{R_{\textrm{global}}^{[c]}\} \le \bar{R}_{\textrm{global}}^{[c]} = (N_C-\alpha) \log\left(1 + \frac{(N_T-\alpha)e^{\frac{N_0}{2}}}{\left(\frac{N_0}{2}\right)^{2-N_C}} \Gamma\left(2-N_C,\frac{N_0}{2}\right)\right),
\end{equation}
\else
\begin{multline} \label{eq:R_global_upper}
    \mathrm{E}\{R_{\textrm{global}}^{[c]}\} \le \bar{R}_{\textrm{global}}^{[c]} = \\ (N_C-\alpha) \log\left(1 + \frac{(N_T-\alpha)e^{\frac{N_0}{2}}}{\left(\frac{N_0}{2}\right)^{2-N_C}} \Gamma\left(2-N_C,\frac{N_0}{2}\right)\right),
\end{multline}
\fi
where $\Gamma(s,t)=\int_{t}^{\infty}x^{s-1} e^{-x}\mathrm{dx}$ is the incomplete gamma function.
\end{lemma}
\begin{IEEEproof}
for $\alpha\leq N_T-1$, the expectation of $R_{\textrm{global}}^{[c]}$ can be bounded as follows:
\begin{align}
    \mathrm{E}\left\{R_{\textrm{global}}^{[c]}\right\} & = \mathrm{E}\left\{\sum_{n\in\mathcal{N}_{C}\setminus\mathcal{F}_{c}}\log\left(1+\eta_{nn}^{[c]}/T_n^{[c]}\right)\right\} \\
    & \leq \sum_{n\in\mathcal{N}_{C}\setminus\mathcal{F}_{c}}\log\left(1+\mathrm{E}\left\{\eta_{nn}^{[c]}\right\}\mathrm{E}\left\{1/T_n^{[c]}\right\}\right). \label{eq:R_global1}
\end{align}

i) $\mathrm{E}\left\{\eta_{nn}^{[c]}\right\}$, $n\in\mathcal{N}_C\setminus\mathcal{F}_c$: For $\alpha=N_T-1$, $\mathbf{w}_{n}^{[c]}$ is designed independently with $\mathbf{h}_{nn}$, and hence, $\eta_{nn}^{[c]}=\left|\mathbf{h}_{nn}^{H}\mathbf{w}_{n}^{[c]}\right|^{2}$ is a Chi-square random variable with degrees of freedom (DoF) 2. On the other hand, for $\alpha\leq N_T-2$, $\mathbf{w}_{n}^{[c]}$ lies in the orthogonal projection of $\mathbf{h}_{nn}$ onto the null space of $\mathbf{h}_{nm}$, $m\in\mathcal{F}_c$. Let us denote $\mathbf{b}_p$ as the $p$-th basis vector of the null space of $\mathbf{h}_{nm}$. The rank of the space composed of these basis vectors is $(N_T-\alpha)$. Then, the desired channel gain can be represented as
\begin{equation}
    \left|\mathbf{h}_{nn}^{H}\mathbf{w}_{n}^{[c]}\right|^2 = \left\Vert\displaystyle\sum_{p=1}^{N_T-\alpha}\left|\mathbf{h}_{nn}\mathbf{b}_{p}\right|\cdot\mathbf{b}_{p}\right\Vert^2 = \displaystyle\sum_{p=1}^{N_T-\alpha}\left|\mathbf{h}_{nn}\mathbf{b}_{p}\right|^2,
\end{equation}
and it is a Chi-square random variable with DoF of $2(N_T-\alpha)$ which is $2$ for $\alpha= N_T-1$.
Thus, we get $\mathrm{E}\left\{\eta_{nn}^{[c]}\right\}=2(N_T-\alpha)$ for $\alpha\leq N_T-1$.

ii) $\mathrm{E}\left\{1/T_n^{[c]}\right\}$, $n\in\mathcal{N}_C\setminus\mathcal{F}_c$: For $\alpha\leq N_T-1$, $\mathbf{w}_{k}^{[c]}$, $k\in\mathcal{N}_{C}\setminus\{n\}$, is designed independently with $\mathbf{h}_{kn}$. Thus, $\eta_{kn}^{[c]}=\left|\mathbf{h}_{kn}^{H}\mathbf{w}_{k}^{[c]}\right|^{2}$ is a Chi-square random variable with DoF 2. Then, we get
\begin{equation}
    \mathrm{E}\left\{1/T_n^{[c]}\right\} = e^{\frac{N_0}{2}}\cdot2^{1-N_C}\cdot N_0^{N_C-2}\cdot\Gamma\left(2-N_C,N_0/2\right),
\end{equation}
where $\Gamma(s)=(s-1)!$ is the gamma function.

From the above two results, the expectation of $R_{\textrm{global}}^{[c]}$ for $\alpha\leq N_T-1$ can be further bounded as follows:
\begin{equation}
    \mathrm{E}\left\{R_{\textrm{global}}^{[c]}\right\} \le (N_C-\alpha) \log\left(1 + \frac{(N_T-\alpha)\Gamma\left(2-N_C,\frac{N_0}{2}\right)}{e^{-\frac{N_0}{2}}\left(\frac{N_0}{2}\right)^{2-N_C}} \right),
\end{equation}
which proves the lemma.
\end{IEEEproof}

From Lemma \ref{lemma:R_global_upper}, we propose to select the index set $\mathcal{F}$ for $\alpha\le N_T-1$, which maximizes $R_{\textrm{local}}^{[c]}+\bar{R}_{\textrm{global}}^{[c]}$. Note that $\bar{R}_{\textrm{global}}^{[c]}=0$ for $\left|\mathcal{F}\right|=\alpha=N_T$, and hence the cost function $R_{\textrm{local}}^{[c]}+\bar{R}_{\textrm{global}}^{[c]}$ can be used for all possible $\alpha$ values discussed.
At this point, to compromise between the amount of information exchange among BSs and the sum-rate performance, let us assume that the information of $r_{m}^{[c]}$, $m\in \mathcal{F}_c$, is collected only for the cases with selected $\alpha$. In this case, let us denote the set of considered $\alpha$ and the index set of the considered cases as $\mathcal{A}$ and $\mathcal{N}_G$, respectively. If the set of considered $\alpha$ is $\mathcal{A}=\{N_T-2, N_T\}$ for $N_T=3$ and $N_C=4$, we have $\left|\mathcal{N}_G\right|={N_C \choose N_T}+{N_C \choose N_T-2}=8$. Then, the index set $\mathcal{F}$ optimization problem is formulated as
\begin{equation} \label{eq:F_opt}
\mathcal{F}=\mathcal{F}_{c^{*}},
\end{equation}
where
\begin{equation}
c^{*}=\arg\max_{c\in\mathcal{N}_G}R_{\textrm{local}}^{[c]}+\bar{R}_{\textrm{global}}^{[c]}.\label{eq:problem}
\end{equation}

\subsubsection{Tightness of the upper bound $\bar{R}_{\textrm{global}}^{[c]}$}
The gap of $\mathrm{E}\{R_{\textrm{global}}^{[c]}\}$ and $\bar{R}_{\textrm{global}}^{[c]}$ results only from the Jensen's inequality in \eqref{eq:R_global1}. The analysis of Jensen's gap has been extensively studied in the literature \cite{Shoshana16,Gao_2018_arXiv}. The gap in the inequality \eqref{eq:R_global1} tends to 0 if the random variable $X_n=1+\eta_{nn}^{[c]}/T_n^{[c]}$ is almost surely constant. The bound of the gap in case where $X_n$ is mean-centric is derived in \cite{Gao_2018_arXiv}. In addition, the $\log$ function becomes an affine function for small $X_n$, resulting in the gap tending to 0. In summary, as received interference at user $n$, $n\in \mathcal{N}_C \setminus\mathcal{F}$, becomes significantly stronger than the desired signal gain, the gap in \eqref{eq:R_global_upper} tends to zero. Furthermore, the more the SINR $X_n$ becomes mean-centric, the tighter upper bound we can get from \eqref{eq:R_global_upper}.

\subsubsection{Asymptotic performance of using $\bar{R}_{\textrm{global}}^{[c]}$}
In the high SNR regime, i.e., $N_{0}$ is arbitrarily small, $R_{\textrm{local}}^{[c]}$ becomes dominant in \eqref{eq:R_c} and we have $R^{[c]}=R_{\textrm{local}}^{[c]}+R_{\textrm{global}}^{[c]}\simeq R_{\textrm{local}}^{[c]}$, for all $c\in\mathcal\{1,\ldots,N_K\}$. In addition, $\bar{R}_{\textrm{global}}^{[c]}$ also tends to 0 in the high SNR regime.  Therefore, the proposed design is asymptotically optimal as SNR increases. In finite SNR regime, as $\alpha$ grows for fixed $N_{C}$, $R_{\textrm{local}}^{[c]}$ in \eqref{eq:R_c} becomes dominant since the number of rate terms in $R_{\textrm{global}}^{[c]}$ which require global CSI, $(N_{C}-\alpha)$, decreases. On the other hand, as $N_{C}$ increases for fixed $\alpha$, the number of interference terms $\eta_{kn}^{[c]}$ in $R_{\textrm{global}}^{[c]}$ increases, and the number of rate terms in $R_{\textrm{global}}^{[c]}$ also increases. It can be readily shown that this global CSI term tends to be bounded by a constant value even in the high-SNR regime, following the analysis in \cite{A_Gupta15_WCL}. Hence, for high-SNR regime, where the $R_{\textrm{local}}^{[c]}$ terms tend to be infinite, or for large $\alpha$ compared with $N_{C}$, the global CSI terms $R_{\textrm{global}}^{[c]}$ become negligible compared to the local CSI terms, resulting in $\bar{R}_{\textrm{global}}^{[c]}$ also tending to 0.

\subsubsection{Performance of using $\bar{R}_{\textrm{global}}^{[c]}$ in finite SNR, $N_T$, and $N_C$}
Figure \ref{fig:R_global_upper} shows the per-cell average $R_{\textrm{global}}^{[c]}$ and $\bar{R}_{\textrm{global}}^{[c]}$ versus SNR for $N_T=4$ and $N_C=7$, where each channel is i.i.d. according to the complex Gaussian distribution. As shown in this figure, the gap between $\textrm{E}\{R_{\textrm{global}}^{[c]}\}$ and $\bar{R}_{\textrm{global}}^{[c]}$ is smaller than 0.04bps/Hz for all possible $\alpha$ values, showing that  $\bar{R}_{\textrm{global}}^{[c]}$ is a good estimator of $\textrm{E}\{R_{\textrm{global}}^{[c]}\}$ even with finite parameter values.
\begin{center}
\begin{figure}[tbh]
\centering{}\includegraphics[width=\if\mycmd1 0.6 \else 0.25 \fi\paperwidth]{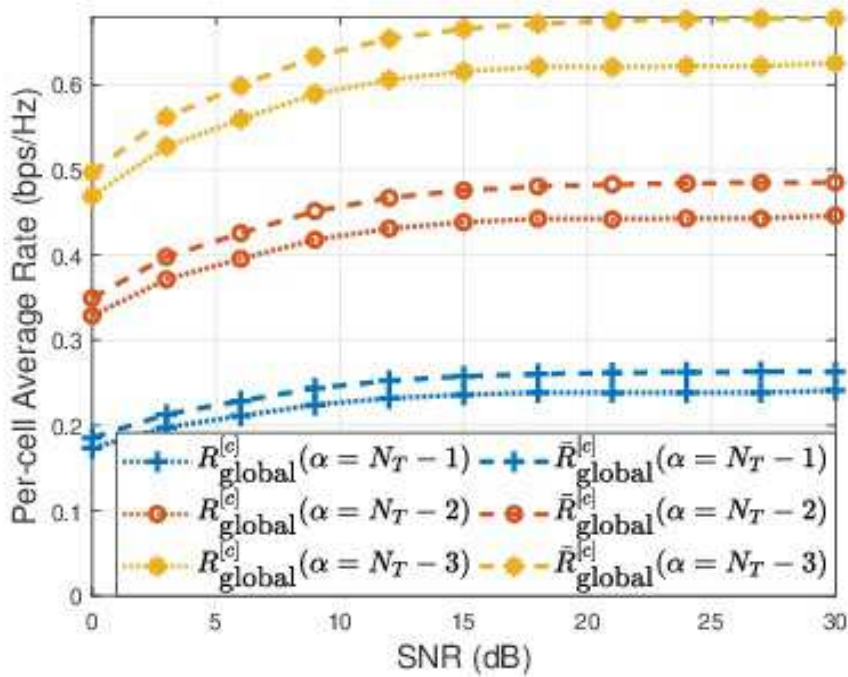}\caption{Per-cell average $R_{\textrm{global}}^{[c]}$ and $\bar{R}_{\textrm{global}}^{[c]}$ versus SNR for $N_T=4$ and $N_C=7$}\label{fig:R_global_upper}
\end{figure}
\par\end{center}

\section{Information Exchange Protocol and Quantization} \label{sec:quantization}
In this section, an information exchange protocol and quantization method are proposed based on the beamforming vector design proposed in Section \ref{sec:precoder}.
\subsection{Information exchange}
To compute the cost function of the problem \eqref{eq:problem}, $R_{\textrm{local}}^{[c]}+\bar{R}_{\textrm{global}}^{[c]}$, each rate term of $R_{\textrm{local}}^{[c]}$, $\log\left( 1+ \eta_{mm}^{[c]}/N_0\right)$, needs to be computed by BS $m$, $m\in\mathcal{F}_c$, with local CSI and be shared by all the BSs. The term $\bar{R}_{\textrm{global}}^{[c]}$ can be computed by any BS without any extra information on instantaneous channels. Let us denote the rate of user $m$ for $m\in\mathcal{F}_c$ with the $c$-th interference-free user selection by
\begin{equation}
    r_{m}^{[c]} = \log\left( 1+ \eta_{mm}^{[c]}/N_0\right). \label{eq:r_ic}
\end{equation}
An example case is as shown in Table \ref{table: rate_table}, where $N_T=3$, $N_C=4$, and $\mathcal{A}=\{N_T-2,N_T\}$. Here, BS 1 can compute the achievable rates in the white cells of the column of BS1 with only local CSI and does not compute the achievable rates correspond to the dark gray cells in the column of BS1 in Table \ref{table: rate_table}, because they require global CSI to computed.
Though each BS can compute ${3 \choose 2}+{3 \choose 1}+{3 \choose 0}=7$ rate terms with local CSI, BS $m$ shares $r_m^{[c]}$ values only for $c\in\mathcal{N}_G$ to restrict the amount of information exchange. Then, for given $c$, $c\in\mathcal{N}_G$, $R_{\textrm{local}}^{[c]}$ is computed by adding all the collected rate terms, i.e., collected rate terms in each row of Table \ref{table: rate_table}, the problem \eqref{eq:problem} can be formulated together with $\bar{R}_{\textrm{global}}^{[c]}$.

\begin{table}
\centering
  \caption{Achievable rates table for all the interference-free user selection cases for $N_{T}=3$ and $N_{C}=4$}\label{table: rate_table}
  \includegraphics[width=\if\mycmd1 0.8 \else 0.4 \fi\paperwidth]{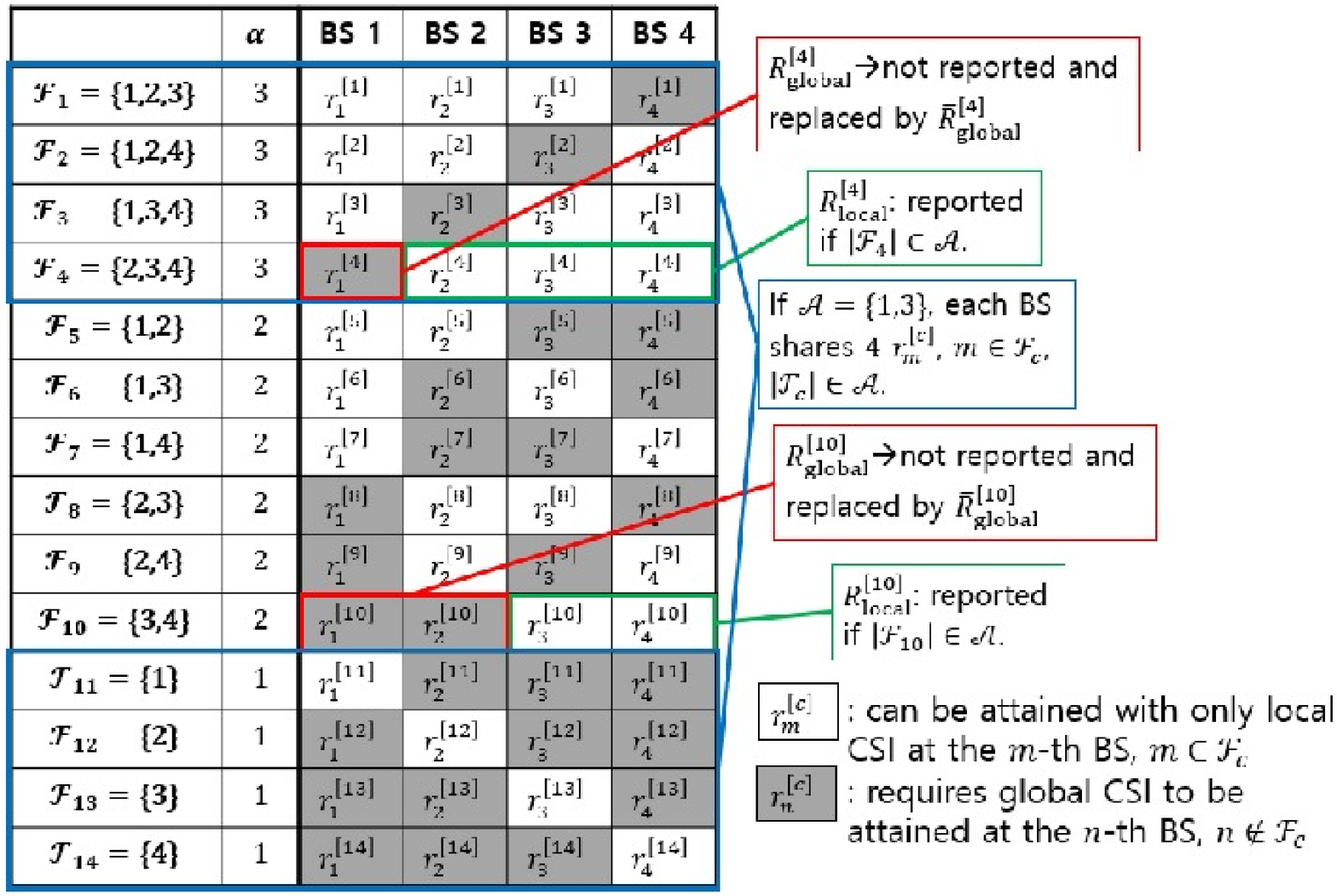}
\end{table}

\subsection{Quantization optimization\label{subsec:Quantization-Optimization}}
In this subsection, the quantization of rate terms that need to be exchanged is analyzed, which is crucial to exchange the information with finite bits.
Let us denote the number of nonzero rates to be exchanged by $M$ and the number of information exchange bits to be used for quantization of each rate by $n_f$. BS $m$ quantizes $M$ rates terms, i.e., $r_{m}^{[c]}$, $c\in\mathcal{N}_G$, $m\in\mathcal{F}_c$. Thus, the number of information exchange bits used at each BS is
\begin{equation} \label{eq:N_f_def}
    N_{f}=M\cdot n_f.
\end{equation}
For optimal quantization, the probability density function (PDF) of $r_{m}^{[c]}$, $c=1,\ldots,N_K$, $m\in\mathcal{F}_c$, is needed, which is denoted by $f(t)$. To get the PDF $f(t)$, we establish the following lemma.

\begin{lemma}\label{lemma:dist_eta}
The random variable $\left|\mathbf{h}_{mm}^{H}\mathbf{w}_{m}^{[c]}\right|^{2}$, $c=1,\ldots,N_K$, $m\in\mathcal{F}_c$, is distributed as a Chi-square random variable with degrees of freedom (DoF) of $2(N_{T}-\alpha+1)$.
\end{lemma}
\begin{IEEEproof}
i) For $\alpha = N_T$, the beamforming vector $\mathbf{w}_{m}^{[c]}$ is designed to only minimize the GI to user $n$, $n\in\mathcal{F}_c\setminus\{m\}$. Thus, $\mathbf{w}_{m}^{[c]}$ is designed independently with the desired channel vector $\mathbf{h}_{mm}$ and $\left|\mathbf{h}_{mm}^{H}\mathbf{w}_{m}^{[c]}\right|^{2}$ is distributed as a Chi-square random variable with DoF 2.

ii) For $\alpha\leq N_T-1$, the beamforming vector $\mathbf{w}_{m}^{[c]}$ is designed to maximize its WSLNR and it has the direction of the orthogonal projection of $\mathbf{h}_{mm}$ onto the null space of $\mathbf{h}_{mn}$, where $m\in\mathcal{F}_{c}$ and $n\in\mathcal{F}_{c}\setminus\{m\}$.
Let us denote $\mathbf{b}_{p}$ is the $p$-th basis vector of the null space of $\mathbf{h}_{mn}$. The number of the basis vector is $N_{T}-(\alpha-1)$. Then, the desired channel gain can be represented as
\begin{equation}
\left|\mathbf{h}_{mm}^{H}\mathbf{w}_{m}^{[c]}\right|^{2} =\left\Vert \sum_{p=1}^{N_{T}-\alpha+1}\left|\mathbf{h}_{mm}\mathbf{b}_{p}\right|\cdot\mathbf{b}_{p}\right\Vert ^{2} = \sum_{p=1}^{N_{T}-\alpha+1}\left|\mathbf{h}_{mm}\mathbf{b}_{p}\right|^{2},
\end{equation}
and it is the Chi-square random variable with DoF of $2(N_{T}-\alpha+1)$.
\end{IEEEproof}

In addition, using Lemma \ref{lemma:dist_eta}, the following theorem is established to derive $f(t)$ which is the PDF of $r_{m}^{[c]}$, $c=1,\ldots,N_K$, $m\in\mathcal{F}_c$.

\begin{theorem}\label{theorem: PDF}
The PDF of $r_{m}^{[c]}$, $c=1,\ldots,N_K$, $m\in\mathcal{F}_c$, is given by
\begin{equation}\label{eq:pdf}
f(t)=\frac{2^{t}N_0^{N_{T}-\alpha+1}\ln2 \left(2^t-1\right)^{N_{T}-\alpha}}{2^{N_{T}-\alpha+1}\Gamma\left(N_{T}-\alpha+1\right)} e^{-\frac{N_0\left(2^t-1\right)}{2}}.
\end{equation}
\end{theorem}

\begin{IEEEproof}
Let us denote the PDF of $\eta_{mm}^{[c]}$ as $h(t)$, the cumulative density function (CDF) of $\eta_{mm}^{[c]}$ as $H(t)$, and the CDF of $r_{m}^{[c]}$ as $F(t)$.
From \eqref{eq:r_ic}, we have $F(t)=H\left(N_0\left(2^t-1\right)\right)$ and
\begin{equation}\label{eq:pdf_r}
f(t)=\ln2\cdot N_0 \cdot2^t\cdot h\left(N_0\left(2^t-1\right)\right).
\end{equation}
Since $h(t)=\frac{t^{N_{T}-\alpha}e^{-\frac{t}{2}}}{2^{N_{T}-\alpha+1}\Gamma(N_{T}-\alpha+1)}$ from the results of Lemma \ref{lemma:dist_eta}, $f(t)$ in \eqref{eq:pdf_r} becomes \eqref{eq:pdf}, which proves the theorem.
\end{IEEEproof}

\begin{figure}
    \centering
    \begin{subfigure}[b]{\if\mycmd1 0.4 \else 0.15 \fi\paperwidth}
        \includegraphics[width=\textwidth]{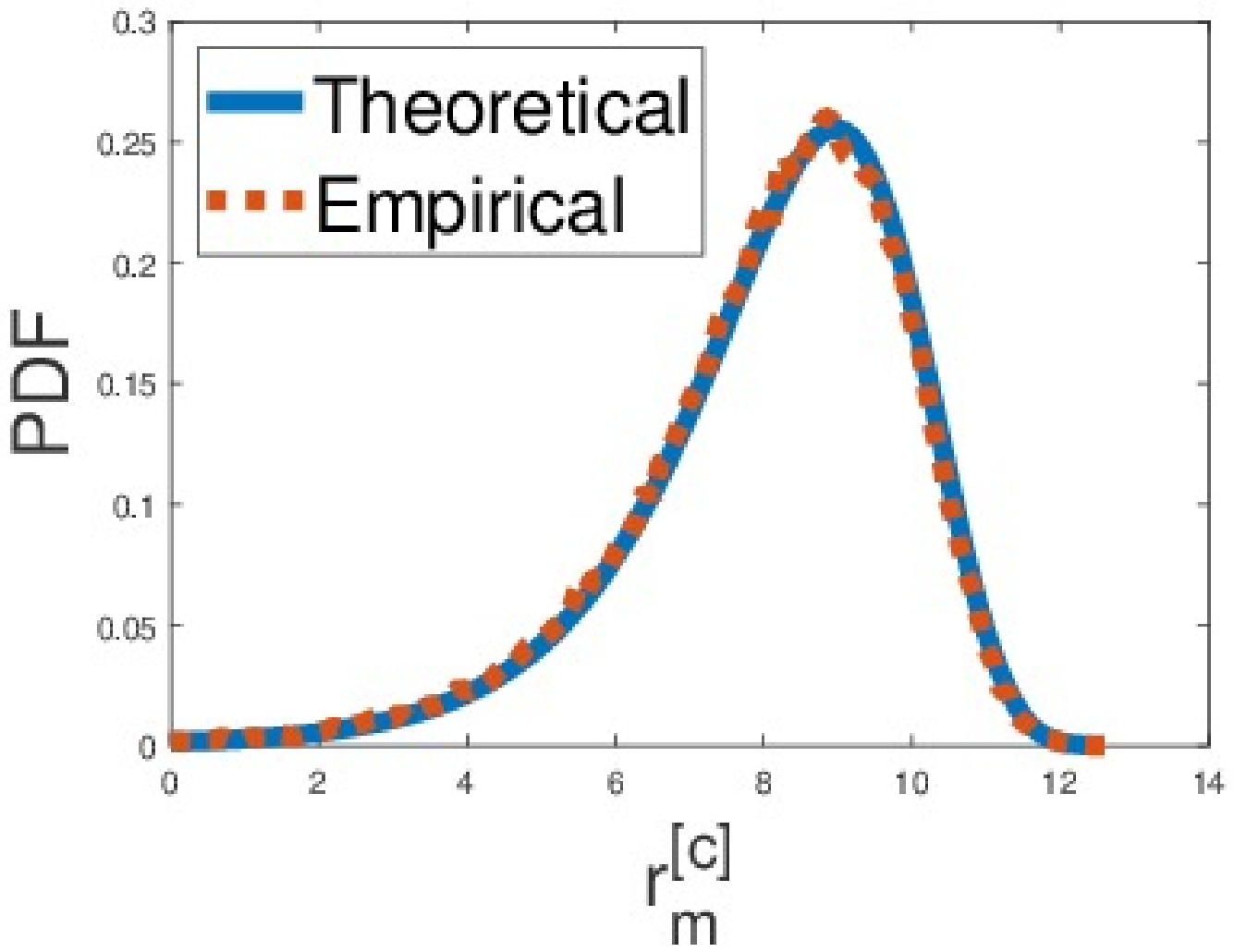}
        \caption{$\alpha=N_T$}
        \label{fig:pdf_Nt}
    \end{subfigure}
        \begin{subfigure}[b]{\if\mycmd1 0.4 \else 0.15 \fi\paperwidth}
        \includegraphics[width=\textwidth]{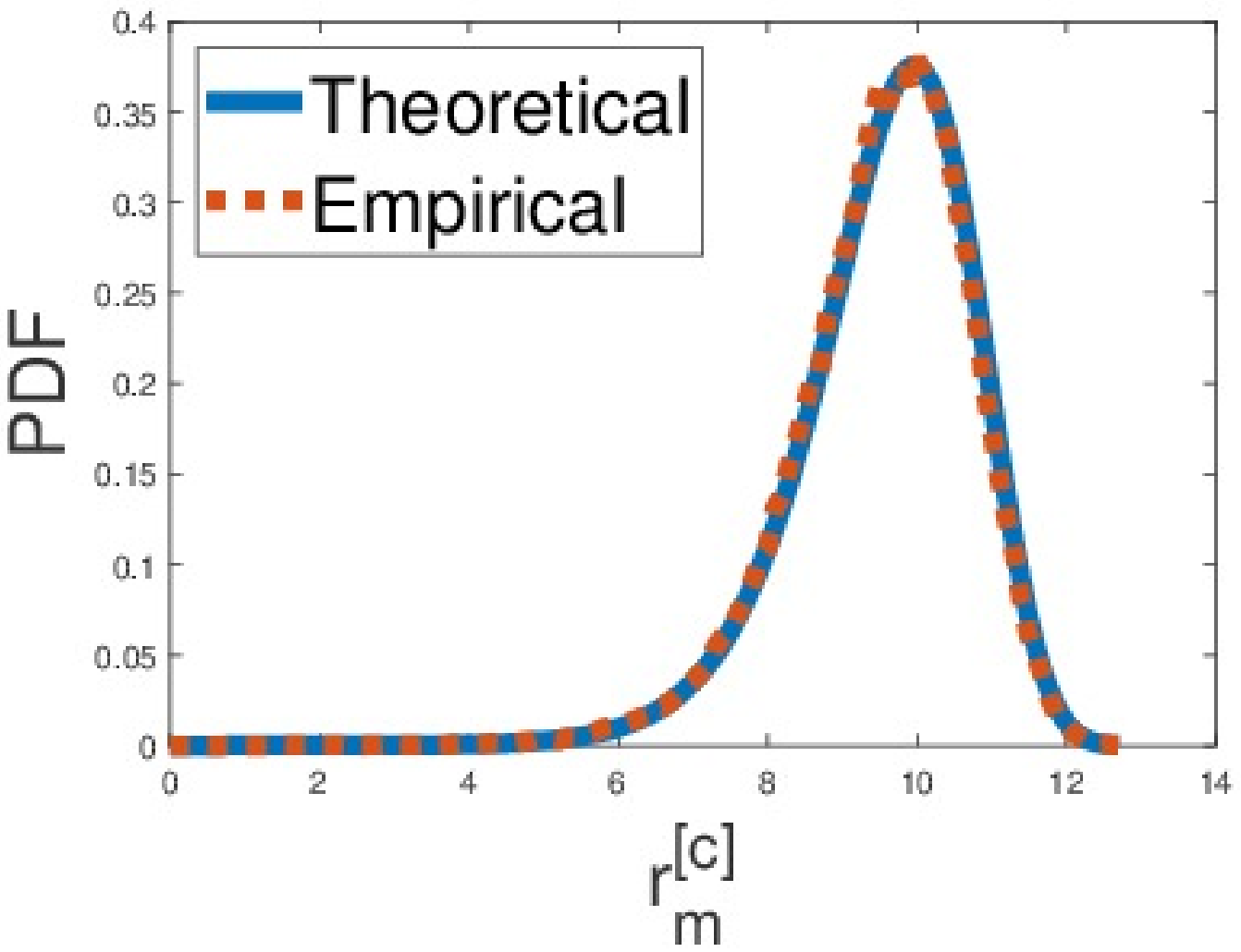}
        \caption{$\alpha = N_T-1$}
        \label{fig:pdf_Nt-1}
    \end{subfigure}
        \begin{subfigure}[b]{\if\mycmd1 0.4 \else 0.15 \fi\paperwidth}
        \includegraphics[width=\textwidth]{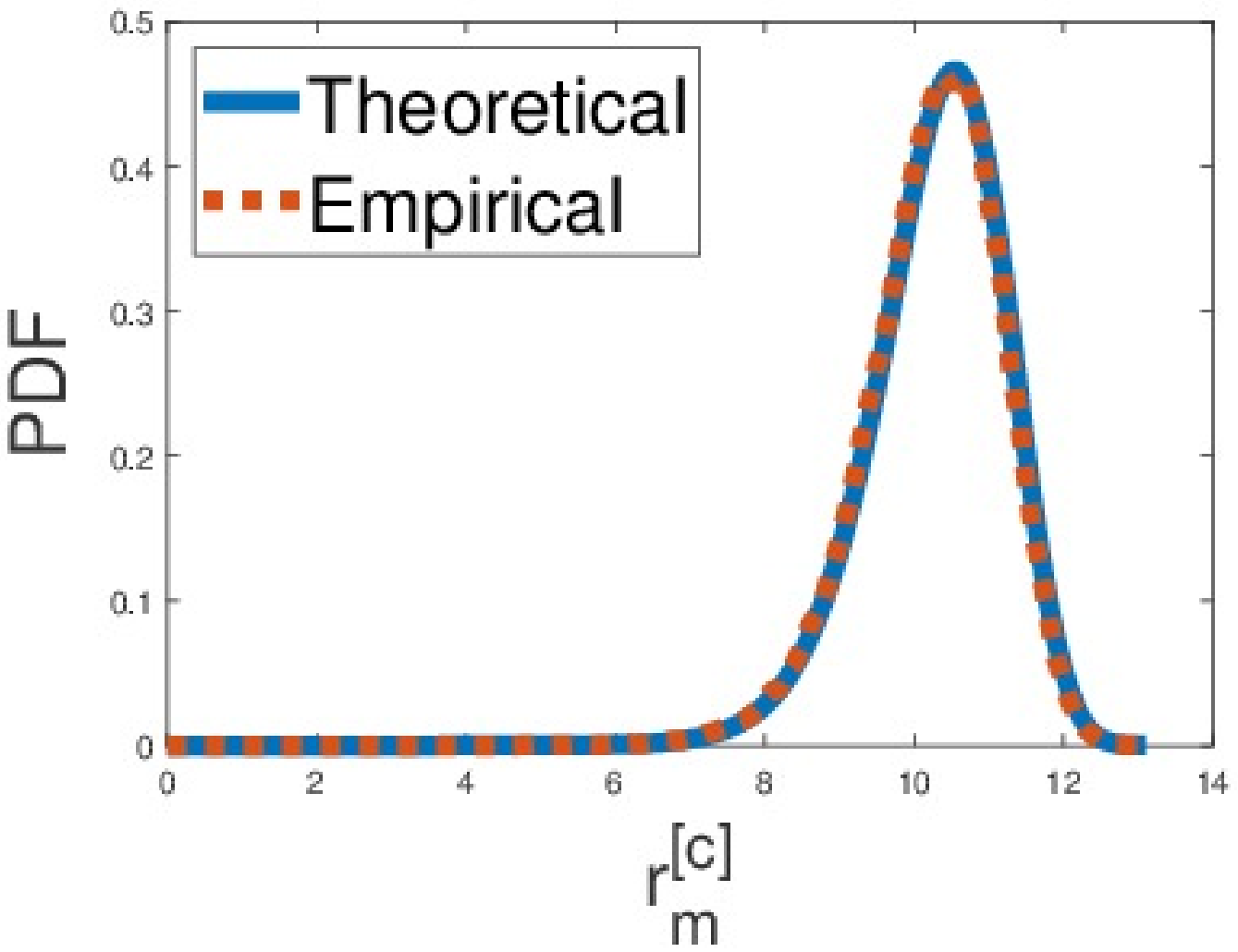}
        \caption{$\alpha = N_T-2$}
        \label{fig:pdf_Nt-2}
    \end{subfigure}
        \caption{PDFs of $r_{m}^{[c]}$ for $N_T=4$, and $N_C=7$}\label{fig:pdf}
\end{figure}

The results in Fig. \ref{fig:pdf} show that the pdf of \eqref{eq:pdf}, denoted by `Theoretical,' is well matched with the simulated histograms which are denoted by `Empirical'.

From the PDF of $r_m^{[c]}$ in Theorem \ref{theorem: PDF}, each $r_m^{[c]}$ is quantized with the Lloyd-max non-uniform quantization method minimizing the mean-square quantization error \cite{S_Lloyd82_TIT}.

\subsection{Information exchange protocol}
\begin{figure}
    \centering
    \begin{subfigure}[b]{\if\mycmd1 0.4  \else 0.25 \fi\paperwidth}
        \includegraphics[width=\textwidth]{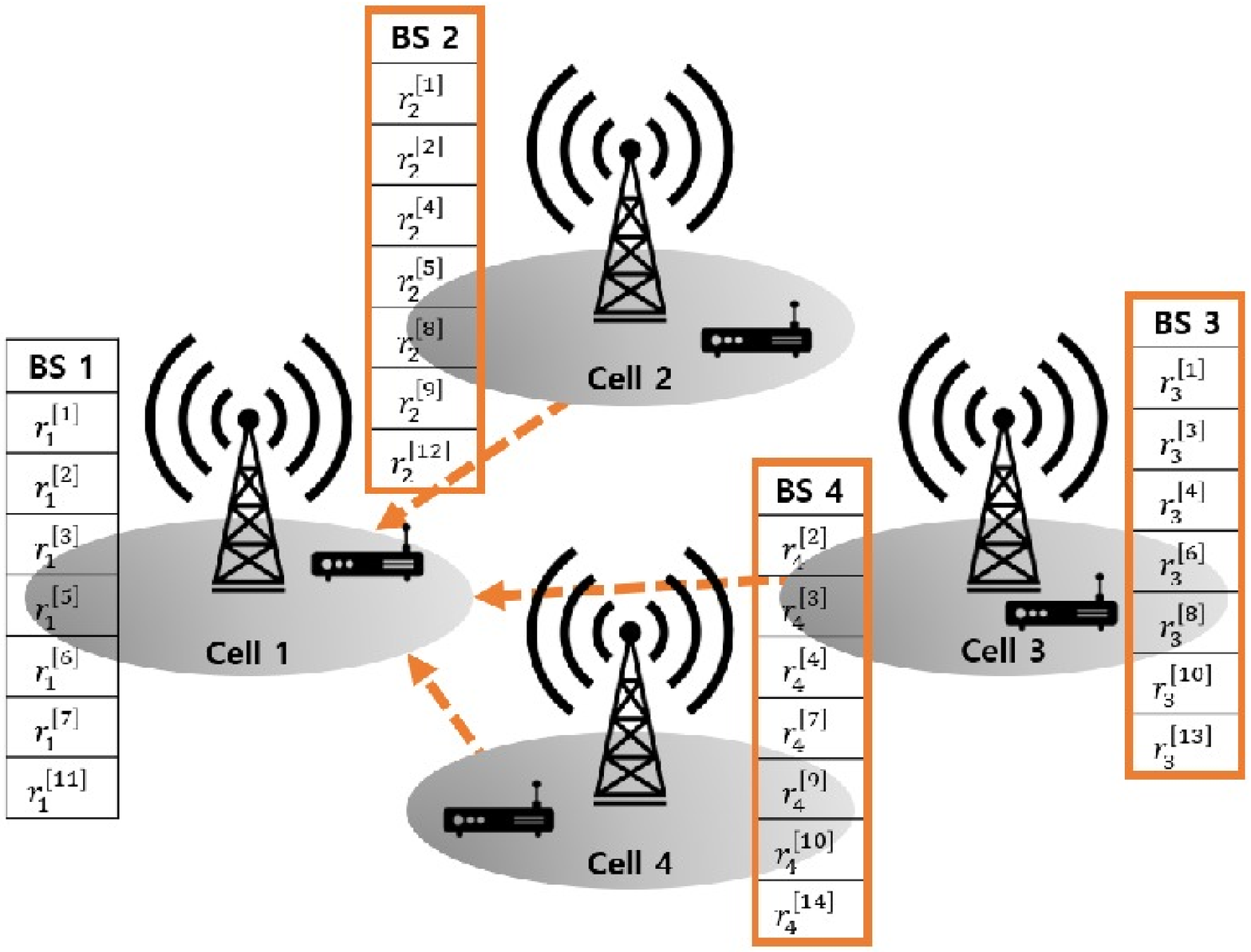}
        \caption{}
        \label{fig:scheme1}
    \end{subfigure} \\
        \begin{subfigure}[b]{\if\mycmd1 0.4  \else 0.3 \fi\paperwidth}
        \includegraphics[width=\textwidth]{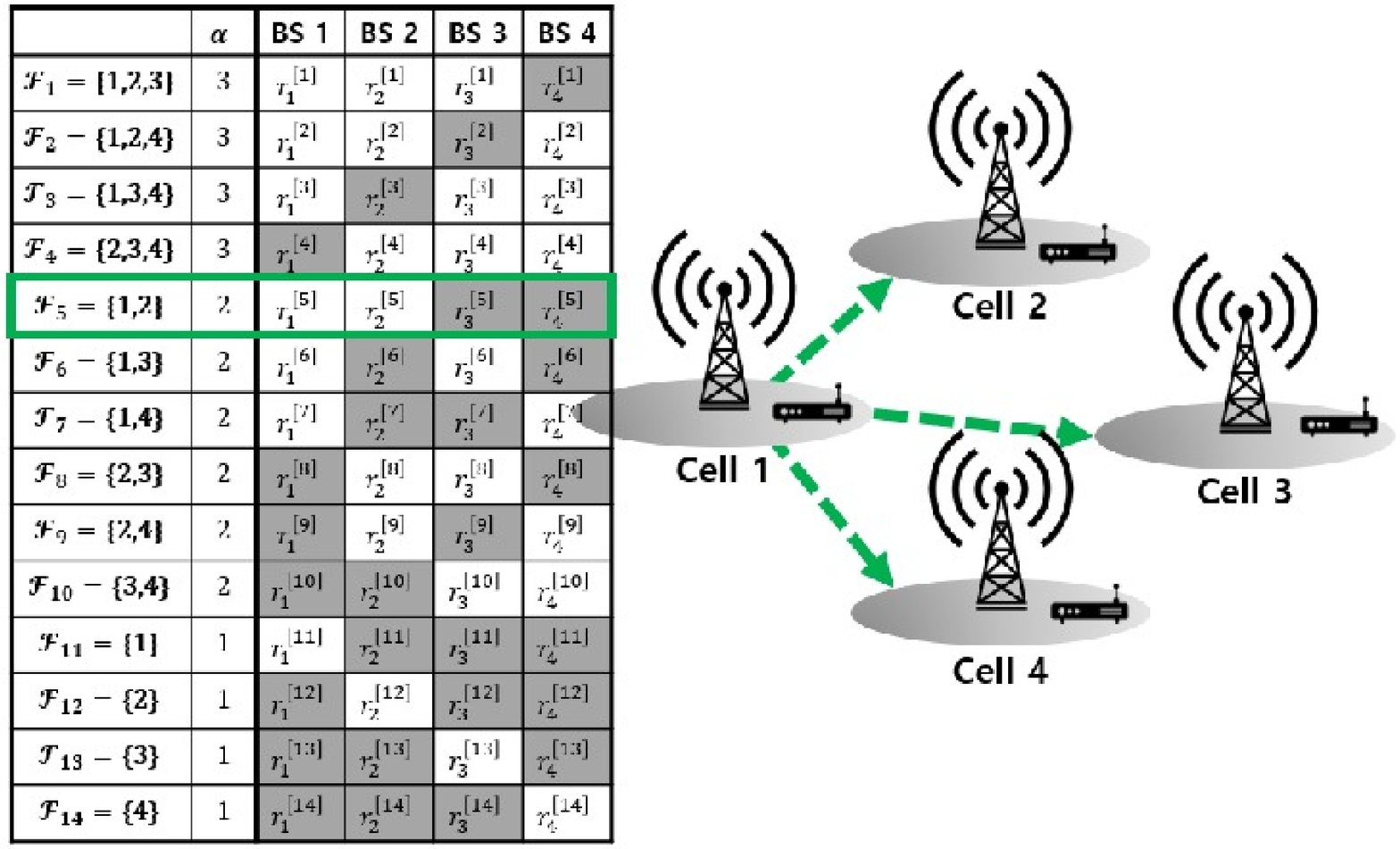}
        \caption{}
        \label{fig:scheme2}
    \end{subfigure} \\
        \begin{subfigure}[b]{\if\mycmd1 0.4  \else 0.25 \fi\paperwidth}
        \includegraphics[width=\textwidth]{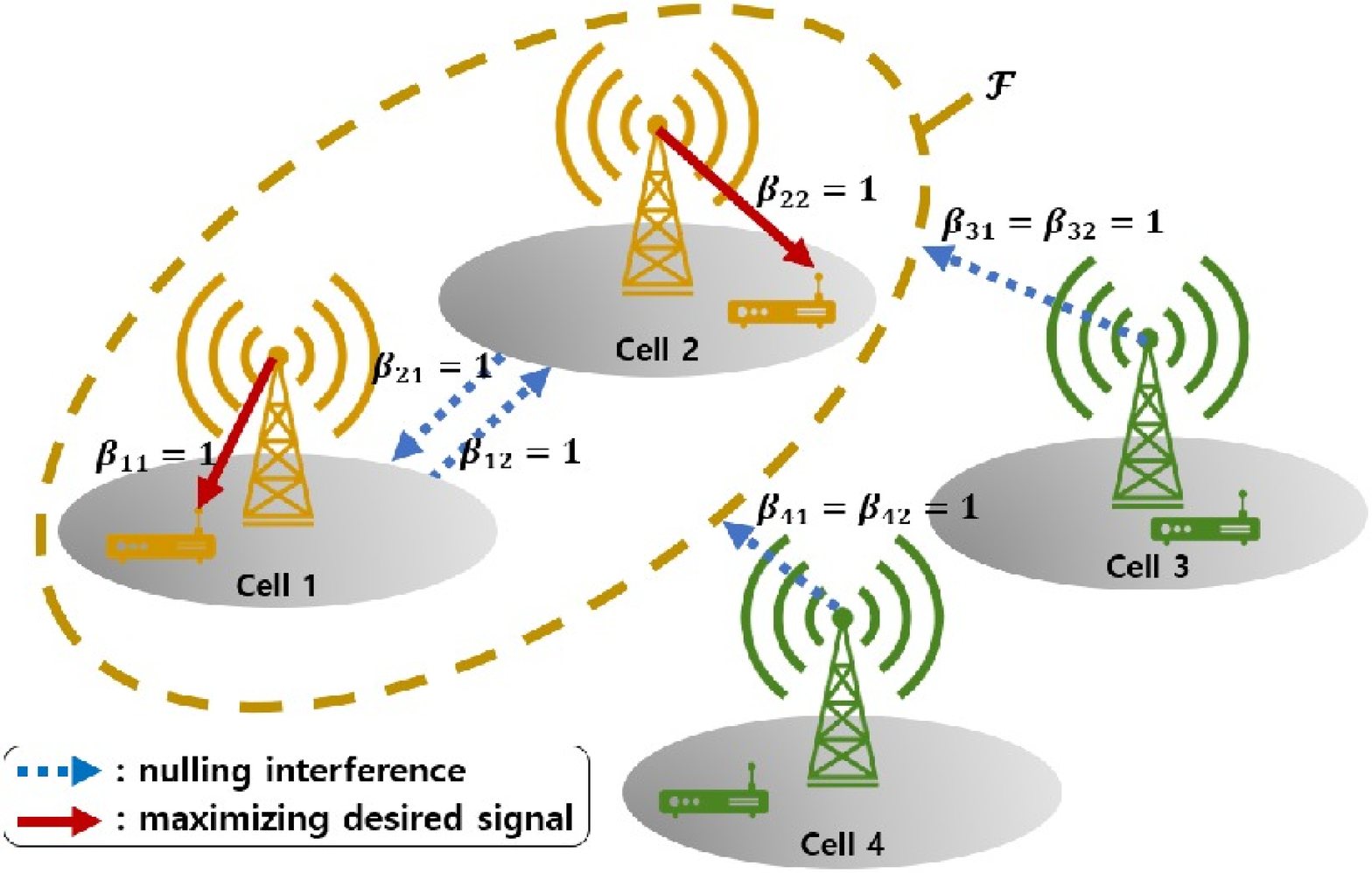}
        \caption{}
        \label{fig:scheme3}
    \end{subfigure}
        \caption{Example of the overall beamforming vectors design and information exchange protocol for $N_T=3$, $N_C=4$, and $\mathcal{A}=\{1,2,3\}$: (a) BS $m$ calculates the rates $r_{m}^{[c]}$ in \eqref{eq:r_ic}, $c\in\mathcal{N}_G$, $m\in\mathcal{F}_c$, which can be calculated with only local CSI and shares them with BS 1 through the information exchange. (b) BS 1 gathers all the $r_{m}^{[c]}$ and makes an table on the left side of Fig. \ref{fig:scheme2}. The white cells in the table are the shared rates by other BSs. Then, BS 1 chooses the set of the interference-free users as \eqref{eq:problem}. The index of the set of the interference-free users is noticed through the information exchange. (c) All the BSs design beamforming vectors which make zero interference to the users with the index which is selected in Fig. \ref{fig:scheme2}. In this example, selected $\alpha$ is $2=N_T-1$ and selected $\mathcal{F}=\{1,2\}$, and hence all the BSs design beamforming vectors as Fig. \ref{fig:beamforming Nt-1}.}\label{fig:scheme}
\end{figure}

There are two possible information exchange protocols. In the first possible protocol, referred to as `centralized protocol,' one of the BSs calculates which users become the interference-free users. A step-by-step illustration of the centralized protocol is depicted in Fig. \ref{fig:scheme}.
In the second possible protocol, referred to as `decentralized protocol,' all the BSs share quantized $M$ rates with the other BSs. Then, each BS determines the interference-free users and designs the beamforming vector.
The total amounts of information exchange bits of the centralized protocol and the decentralized protocol are denoted by $S_{\textrm{centralized}}$ and $S_{\textrm{decentralized}}$, respectively. Then, we have
\begin{equation}
S_{\textrm{central}} =(N_{C}-1)\cdot N_{f}+(N_{C}-1)\cdot\left\lceil\log\left(\sum_{\alpha = 1}^{N_T} {N_{C}\choose\alpha} \right)\right\rceil,
\end{equation}
\begin{equation}
S_{\textrm{decentral}} =(N_{C}-1)\cdot N_{f}+(N_{C}-1)\cdot(N_{C}-1)\cdot N_{f}.
\end{equation}
Since $N_C \geq 3$ and $\sum_{\alpha = 1}^{N_T} {N_{C}\choose\alpha}<N_T\sqrt{\frac{8}{\pi N_{C}}}\cdot 2^{N_{C}-1}$ from \cite[Theorem 2.2]{Z_Sun13_arXiv}, we have
\begin{align}
\left\lceil\log\left(\sum_{\alpha = 1}^{N_T} {N_{C}\choose\alpha}\right)\right\rceil & \leq(N_{C}-1)+\left\lceil\log N_T\right\rceil \label{eq:inequality_S} \\
& < (N_C-1) + N_T \label{eq:inequality_S2}  \\
& \leq (N_C-1)\cdot 2
\end{align}
for $N_f\geq 2$. Thus, the centralized protocol is more preferable than the decentralized protocol, especially if $N_f \geq 2$, and hence we use the centralized protocol as shown in Fig. \ref{fig:scheme}.

\subsection{Information exchange comparison} \label{sec:information exchange comparison}

In this subsection, the amount of information exchange required in the proposed scheme is compared to those of existing schemes.
The distributed weighted minimizing mean-square error (WMMSE) scheme \cite{Q_Shi11_TSP} is considered, where each beamforming vector is designed iteratively between the transmitters and receivers. It is known that the `WMMSE' scheme is the most efficient scheme that iteratively achieves the optimal sum-rate bound but in a distributed manner.
The number of iteration and the number of bits required for the quantization of each scalar or vector in the `WMMSE' scheme are denoted by $\kappa$ and $n_f^{\textrm{WMMSE}}$, respectively.
The `Global' scheme  is also considered, where all the beamforming vectors are jointly optimized in pursuit of maximizing the sum-rate with global CSI \cite{R_Bhagavatula10_ICASSP}. Let us denote the number of bits required for the quantization of each vector in the `Global' scheme by $n_f^{\textrm{Global}}$.
Then, Table \ref{table: backhaul_signaling} summarizes the amount of required information exchange in bytes for the considered schemes.
As shown in Table \ref{table: backhaul_signaling}, the amount of the required information exchange of the propose scheme is much less than those of `WMMSE' and `Global'.
Moreover, the required information exchange of `WMMSE' increases in proportion to the number of iteration $\kappa$.
Unlike `WMMSE,' the information required to be exchanged among BSs is merely scalar values. Therefore, the amount of information exchange does not increase even for growing $N_T$, which significantly lowers the burden of the backhaul or direct link.

\begin{table*}[tbh]
\caption{Amount of required information exchange}\label{table: backhaul_signaling}
\begin{center}
\begin{tabular}{|c|c|c|c|}
\hline
\multirow{2}{*}{Scheme} & \multicolumn{3}{c|}{Amount of information exchange (bytes)}                                                                                                                                                                                                          \\ \cline{2-4}
                        & General expression (bits)                                                                   & \begin{tabular}[c]{@{}c@{}}$N_T=4$, $N_C=7$, $N_f=35$,\\ $n_f^{\textrm{WMMSE}}=n_f^{\textrm{Global}}=2$, $\kappa=2$\end{tabular} & \begin{tabular}[c]{@{}c@{}}$N_T=8$, $N_C=9$, $N_f=84$,\\ $n_f^{\textrm{WMMSE}}=n_f^{\textrm{Global}}=5$, $\kappa=2$\end{tabular} \\ \hline
Proposed                & $(N_C-1)\cdot\left(N_f+\left\lceil\log(\sum_{\alpha = 1}^{N_T} {N_{C}\choose\alpha} )\right\rceil\right)$ & 32                                                                                        & 91                                                                                      \\ \hline
WMMSE                   & $3\kappa n_f^{\textrm{WMMSE}} N_{C}^2$                                                      & 74                                                                                        & 304                                                                                      \\ \hline
Global                 & $n_f^{\textrm{Global}} N_{C}^2 (N_C-1)$                                                      & 74                                                                                        & 405                                                                                      \\ \hline
\end{tabular}
\end{center}
\end{table*}

\section{Extension to the Multiuser Case}
\label{sec:multiuser}

In this section, the proposed scheme is extended to the multiuser case, where each cell is composed of $N_U$ users. User $p$ in the $i$-th cell is referred to as user $i_p$, where $i\in\mathcal{N}_{C}$, $p\in\{1,\ldots,N_U\}\triangleq\mathcal{N}_{U}$, and $i_p\in\{i_p|i\in\mathcal{N}_C,p\in\mathcal{N}_U\}\triangleq\mathcal{N}_{W}$. It is assumed that $N_T<N_U N_C$. The channel vector from BS $i$ to user $j_r$ is denoted by $\mathbf{h}_{i,j_r}$.
The received signal at user $i_p$ is written by
\if\mycmd1
\begin{equation}
    y_{i_p}=\underbrace{\mathbf{h}_{i,i_p}\mathbf{w}_{i_p}x_{i_p}}_{\textrm{desired\ signal}} + \underbrace{\sum_{q\in\mathcal{N}_C\setminus\{p\}}\mathbf{h}_{i,i_p}\mathbf{w}_{i_q}x_{i_q}}_{\textrm{intracell\ interference}} +\underbrace{\sum_{j\in\mathcal{N}_C\setminus\{i\}}\sum_{r\in\mathcal{N}_U}\mathbf{h}_{j,i_p}\mathbf{w}_{j_r}x_{j_r}}_{\textrm{intercell\ interference}} + z_{i_p},
\end{equation}
\else
\begin{multline}
y_{i_p}=\underbrace{\mathbf{h}_{i,i_p}\mathbf{w}_{i_p}x_{i_p}}_{\textrm{desired\ signal}} + \underbrace{\sum_{q\in\mathcal{N}_C\setminus\{p\}}\mathbf{h}_{i,i_p}\mathbf{w}_{i_q}x_{i_q}}_{\textrm{intracell\ interference}} \\
+\underbrace{\sum_{j\in\mathcal{N}_C\setminus\{i\}}\sum_{r\in\mathcal{N}_U}\mathbf{h}_{j,i_p}\mathbf{w}_{j_r}x_{j_r}}_{\textrm{intercell\ interference}} + z_{i_p},
\end{multline}
\fi
where $x_{l_k}$ is the unit-variance transmit symbol at the $l$-th
BS to user $l_k$ and $z_{l_k}$ is the AWGN at user $l_k$ with zero-mean and variance of $N_{0}$. Thus, the corresponding SINR is expressed by
\begin{equation}
\gamma_{i_p}=\frac{|\mathbf{h}_{i,i_p}\mathbf{w}_{i_p}|^2}{\displaystyle\sum_{q\in\mathcal{N}_U\setminus\{p\}}|\mathbf{h}_{i,i_p}\mathbf{w}_{i_q}|^2+\displaystyle\sum_{j\in\mathcal{N}_C\setminus\{i\}}\displaystyle\sum_{r\in\mathcal{N}_U}|\mathbf{h}_{j,i_p}\mathbf{w}_{j_r}|^2+N_0},
\end{equation}
and the achievable sum-rate is given by
\begin{equation}
R_M=\sum_{i\in\mathcal{N}_C}\sum_{p\in\mathcal{N}_U}\log(1+\gamma_{i_p}).
\end{equation}

As shown in Section \ref{sec:precoder_A}, the sum-rate maximization problem can be obtained by solving the max-WSLNR problem. Let us denote the weight coefficient for the channel gain from BS $i$ to user $j_r$ by $\beta_{i,j_r}\geq0$ and the set of $\beta_{i,j_r}$, $j_r\in\mathcal{N}_{W}$, by $\boldsymbol{\beta}_{i}=\{\beta_{i,j_r}|j_r\in\mathcal{N}_{W}\}$. As Section \ref{sec:precoder}, $\beta_{i,j_r}$ is restricted to $\beta_{i,j_r}\in\{0,1\}$ and then a general WSLNR incorporating the notion of WGI is defined as follows:
\begin{equation}\label{eq:WSLNR_multi}
    \chi_{i_p} =
    \begin{cases}
    \beta_{i,i_p}|\mathbf{h}_{i,i_p}\mathbf{w}_{i_p}|^2/\left(P_{i_p}+N_0\right) & \text{ if } \beta_{i,i_p}\neq0\\
    1/\left(P_{i_p}+N_0\right) & \text{ if } \beta_{i,i_p}=0,
    \end{cases}
\end{equation}
where
\begin{equation}
    P_{i_p} = \displaystyle\sum_{q\in\mathcal{N}_U\setminus\{p\}}\beta_{i,i_q}|\mathbf{h}_{i,i_q}\mathbf{w}_{i_p}|^2+\displaystyle\sum_{j\in\mathcal{N}_C\setminus\{i\}}\displaystyle\sum_{r\in\mathcal{N}_U}\beta_{i,j_r}|\mathbf{h}_{i,j_r}\mathbf{w}_{i_p}|^2.
\end{equation}
In this section, the set of all the interference-free users is denoted by $\mathcal{F}_M\subset\mathcal{N}_{W}$, and the number of interference-free users as $\alpha_{M}$.

\subsection{Beamforming vector design for $|\mathcal{F}_M|=N_T$}
Suppose that user $m_p$ is an interference-free user. Then, the interference-free constraints at the receiver side are given by
\begin{equation}\label{eq:interf_eq2}
\mathbf{h}_{k,m_p}^{H}\mathbf{w}_{k_l}=0,\,\,k_l\in\mathcal{N}_{W}\setminus\{m_p\}.
\end{equation}
The number of these equalities for the $\alpha_{M}$ interference-free users is $\alpha_{M}((N_U-1)+(N_{A}-1)N_U)$ assuming $N_A$ is the number of BSs with non-zero transmission power. On the other hand, the number of effective variables in each $\mathbf{w}_{k_l}$ is $N_{T}-1$ considering the unit-norm constraint. For the existence of the solution on $\mathbf{w}_{k_l}$ of the equalities \eqref{eq:interf_eq2}, we need the number of effective variables to be equal to or greater than the number of equalities, i.e., $\alpha_{M}((N_U-1)+(N_{A}-1)N_U) \le N_{A}N_U(N_{T}-1)\Longleftrightarrow\alpha_{M}\le\frac{N_A N_{U}(N_{T}-1)}{N_{A}N_U-1}$.
Therefore, the maximum number of interference-free users in the MISO interference channel for multiuser case is given by
\begin{equation}\label{eq:alpha_M max}
\alpha_{M,\max} =
\begin{cases}
 N_T & \text{ if } N_A = N_T/N_U \\
 N_T-1 & \text{ if } N_A > N_T/N_U \\
 N_A N_U & \textrm{ otherwise.}
\end{cases}
\end{equation}
As shown in \eqref{eq:alpha_M max}, $|\mathcal{F}_M|=N_T$ can be achieved only when $N_T$ is divisible by $N_U$. If $\frac{N_T}{N_U}$ is a natural number, there exist ${N_C \choose \frac{N_T}{N_U}}$ possible interference-free users selection for $|\mathcal{F}_M|=N_T$.

The $N_T$ interference-free users can be obtained by muting $(N_C-\frac{N_T}{N_U})$ BSs. In such a case, all the users in the cells where the BSs have non-zero transmission power are the interference-free users. BS $m$ which has non-zero transmission power designs beamforming vectors that maximize \eqref{eq:WSLNR_multi} setting $\beta_{m,m_p}=0$, $\beta_{m,k_l}=1$ for $k_l\in\mathcal{F}_M\setminus\{m_p\}$, and $\beta_{m,n_r}=1$ for $n_r\in\mathcal{N}_{W}\setminus\mathcal{F}_M$ as
\begin{align}
    \mathbf{w}_{m_p}^{\textrm{min-WGI}} & = \arg \max_{\left\|\mathbf{w}\right\|^2=1} \frac{1}{\sum_{k_l\in\mathcal{F}_M\setminus\{m_p\}}\left|\mathbf{h}_{m,k_l}^{H}\mathbf{w}\right|^2+N_0}\label{eq: min-WGI-multiuser} \\
    & = \arg \min_{\left\|\mathbf{w}\right\|^2=1} \left\Vert \mathbf{G}_{m_p}\mathbf{w}\right\Vert ^{2},
\end{align}
where $\mathbf{G}_{m_p} \triangleq \left[\sqrt{\beta_{m,1_1}}\mathbf{h}_{m,1_1}, \ldots, \sqrt{\beta_{m,{N_C}_{N_U}}}\mathbf{h}_{m,{N_C}_{N_U}} \right]^{H}\in\mathbb{C}^{N_{C}N_{U}\times N_{T}}$.
Then, the solution of \eqref{eq: min-WGI-multiuser} is obtained by choosing the right singular vector of $\mathbf{G}_{m_p}$ associated with the smallest singular value.

\subsection{Beamforming vector design for $|\mathcal{F}_M|= N_T-1$}
For $|\mathcal{F}_M|= N_T-1$, all the BSs have non-zero transmission power.

\subsubsection{Design of $\mathbf{w}_{n_r}$ for $n_r\in\mathcal{N}_{W}\setminus\mathcal{F}_M$}
BS $n$ designs beamforming vector $\mathbf{w}_{n_r}$, $n_r\in\mathcal{N}_{W}\setminus\mathcal{F}_M$, to make user $m_p$, $m_p\in\mathcal{F}_M$, interference-free. Thus, the beamforming vector design of $\mathbf{w}_{n_r}$ for $n_r\in\mathcal{N}_{W}\setminus\mathcal{F}_M$ employ the min-WGI beamforming design in \eqref{eq: min-WGI-multiuser}.

\subsubsection{Design of $\mathbf{w}_{m_p}$ for $m_p\in\mathcal{F}$}
Since BS $m$ designs the beamfomring vector $\mathbf{w}_{m_p}$, $m_p\in\mathcal{F}$, only making zero interference to the users with the indices in $\mathcal{F}\setminus\{m_p\}$, where $|\mathcal{F}\setminus\{m_p\}|=N_T-2$, BS $m$ utilizes the space of rank one to improve the channel gain. Then, $\mathbf{w}_{m_p}$ is designed maximizing \eqref{eq:WSLNR_multi} with $\beta_{m,m_p}=\beta_{m,k_l}=1$ for $k_l\in\mathcal{F}\setminus\{m_p\}$ and $\beta_{m,n_r}=0$ for $n_r\in\mathcal{N}_{W}\setminus\mathcal{F}$.

\subsection{Beamforming vector design for $|\mathcal{F}_M|\leq N_T-2$}
For $|\mathcal{F}_M|=\alpha_{M}\leq N_T-2$, all the BSs designs the beamforming vectors making zero interference to user $m_p$, $m_p\in\mathcal{F}_M$. The number of neighboring users, to which each BS makes GI zero, is $\alpha_{M}-1$ for the beamforming vector design of $\mathbf{w}_{m_p}$, $m_p\in\mathcal{F}_M$, and $\alpha_{M}$ for the beamforming vector design of $\mathbf{w}_{n_r}$, $n_r\in\mathcal{N}_{W}\setminus\mathcal{F}_M$. Then, the beamforming vectors $\mathbf{w}_{m_p}$ and $\mathbf{w}_{n_r}$ are designed in the null space of ranks $(N_T-\alpha_{M}+1)$ and $(N_T-\alpha_{M})$, respectively. Thus, BS $m$ designs the beamforming vectors $\mathbf{w}_{m_p}$ maximizing \eqref{eq:WSLNR_multi} by setting $\beta_{m,m_p}=\beta_{m,k_l}=1$, $m_p\in\mathcal{F}_M$, $k_l\in\mathcal{F}_M\setminus\{m_p\}$, and $\beta_{m,n_r}=0$, $n_r\in\mathcal{N}_{W}\setminus\mathcal{F}_M$. BS $n$ designs the beamforming vectors $\mathbf{w}_{n_r}$ maximizing \eqref{eq:WSLNR_multi} by setting $\beta_{n,m_p}=\beta_{n,n_r}=1$, $m_p\in\mathcal{F}_M$, $n_r\in\mathcal{N}_{W}\setminus\mathcal{F}_M$, and $\beta_{n,v_g}=0$, $v_g\in\mathcal{N}_{W}\setminus(\mathcal{F}_M\cup\{n_r\})$.

Since BS $m$ designs the beamforming vector $\mathbf{w}_{m_p}$, $m_p\in\mathcal{F}_M$, only making zero interference to the users with the indices in $\mathcal{F}_M\setminus\{m_p\}$, where $|\mathcal{F}_M\setminus\{m_p\}|=\alpha_{M}-1$, BS $m$ utilizes the space of rank $(N_T-\alpha_{M}+1)$ to improve the channel gain. Then, $\mathbf{w}_{m_p}$ is designed maximizing \eqref{eq:WSLNR_multi} with $\beta_{m,m_p}=\beta_{m,k_l}=1$ for $k_l\in\mathcal{F}_M\setminus\{m_p\}$ and $\beta_{m,n_r}=0$ for $n_r\in\mathcal{N}_{W}\setminus\mathcal{F}_M$.

\subsection{Selection of $\mathcal{F}_M$: Design of $\beta_{i,k_l}$}
There exist ${N_C \choose \frac{N_T}{N_U}} + {N_C N_U \choose N_T-1} + \cdots + {N_C N_U \choose 1}$ possible interference-free users selection if $N_T$ is divisible  by $N_U$ and ${N_C N_U \choose N_T-1} + \cdots + {N_C N_U \choose 1}$ possible interference-free users selection otherwise. Let us denote the $c$-th interference-free user selection as $\mathcal{F}_c^{[M]}$. Then, the sum-rate for the $c$-th interference-free users selection can be represented as
\begin{equation}\label{eq:R_c_multiuser}
R^{[c]}_{M} =R_{\textrm{local},M}^{[c]} + R_{\textrm{global},M}^{[c]},
\end{equation}
where $R_{\textrm{local},M}^{[c]}=\sum_{m_p\in\mathcal{F}_c^{[M]}}\log\left(1+\frac{\eta_{m_p,m_p}^{[c]}}{N_{0}}\right)$, $R_{\textrm{global},M}^{[c]}=\sum_{n_r\in\mathcal{N}_{W}\setminus\mathcal{F}_c^{[M]}}\log\left(1+\frac{\eta_{n_r,n_r}^{[c]}}{T_{n_r}^{[c]}}\right)$, $\eta_{i_p,j_q}^{[c]}=|\mathbf{h}_{j,i_p}\mathbf{w}_{j_q}^{[c]}|^2$, and $T_{n_r}^{[c]}=\displaystyle\sum_{s\in\mathcal{N}_U\setminus\{r\}}\eta_{n_r,n_s}^{[c]} + \displaystyle\sum_{k\in\mathcal{N}_C\setminus\{n\}}\displaystyle\sum_{l\in\mathcal{N}_U}\eta_{n_r,k_l}^{[c]}+N_0$.

We propose to select the index set $\mathcal{F}$ which maximizes $R_{\textrm{local},M}^{[c]}+\bar{R}_{\textrm{global},M}^{[c]}$, where
\if\mycmd1
\begin{equation}
\bar{R}_{\textrm{global},M}^{[c]}=(N_C N_U-\alpha_{M})\log\left(1 + \frac{(N_T-\alpha_{M})\Gamma\left(2-N_C N_U,\frac{N_0}{2}\right)}{e^{-\frac{N_0}{2}}\left(\frac{N_0}{2}\right)^{2-N_C N_U}}\right)
\end{equation}
\else
\begin{multline}
\bar{R}_{\textrm{global},M}^{[c]}=(N_C N_U-\alpha_{M}) \\ \times \log\left(1 + \frac{(N_T-\alpha_{M})\Gamma\left(2-N_C N_U,\frac{N_0}{2}\right)}{e^{-\frac{N_0}{2}}\left(\frac{N_0}{2}\right)^{2-N_C N_U}}\right)
\end{multline}
\fi
is the upper bound of $E\{R_{\textrm{global},M}^{[c]}\}$, which can be obtained from Lemma \ref{lemma:R_global_upper} by considering  both intercell interference and intracell interference.
At this point, let us assume that the information is collected only for the cases with selected $\alpha_M$ as in Section \ref{sec:design set of interference-free users}. In   case, let us denote the set of considered $\alpha_M$ and the index set of the considered cases as $\mathcal{A}_M$ and $\mathcal{N}_G^{M}$, respectively. For example, if the set of considered $\alpha_{M}$ is $\mathcal{A}_M=\{N_T-1,N_T\}$ for $N_T=3$, $N_C=4$, and $N_U=3$, we have $|\mathcal{N}_G^M|={N_C \choose \frac{N_T}{N_U}}+{N_C\cdot N_U \choose {N_T-1}}=70$.

Finally, the index set $\mathcal{F}_M$ can be found from
\begin{equation}
    \mathcal{F}_M=\mathcal{F}_{c^*}^{[M]}
\end{equation}
\begin{equation}
    c^{*}=\arg\max_{c\in\mathcal{N}_G}R_{\textrm{local},M}^{[c]}+\bar{R}_{\textrm{global},M}^{[c]}.
\end{equation}

\section{Numerical Simulations}

\label{sec:numerical}
\begin{figure}
    \centering
    \begin{subfigure}[b]{\if\mycmd1 0.6 \else 0.21 \fi\paperwidth}
        \includegraphics[width=\textwidth]{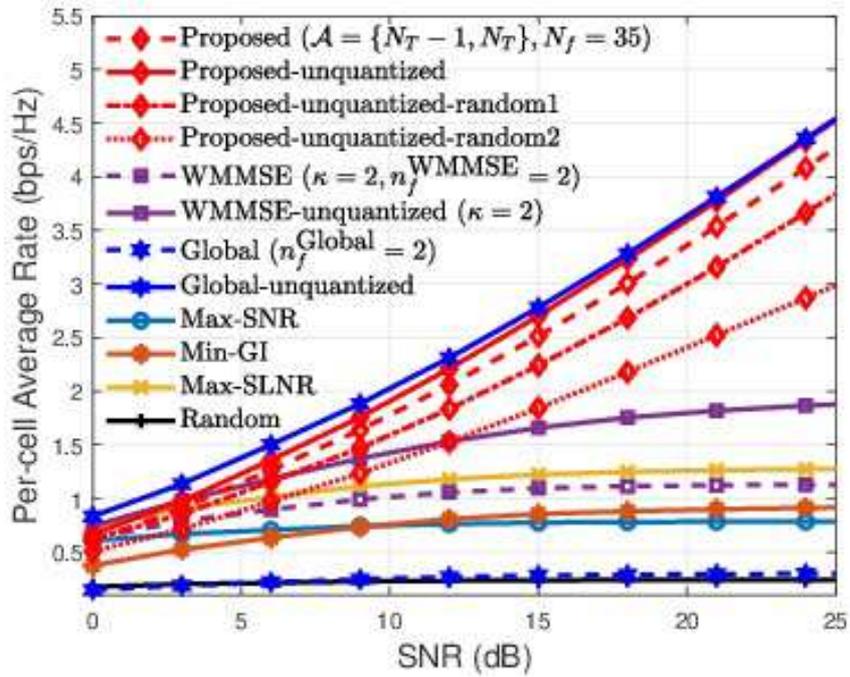}
        \caption{Per-cell average rate versus SNR of the proposed scheme compared to the existing schemes}
        \label{fig:s1_Nt4Nc7_SumRate vs. SNR}
    \end{subfigure}
        \begin{subfigure}[b]{\if\mycmd1 0.6 \else 0.21 \fi\paperwidth}
        \includegraphics[width=\textwidth]{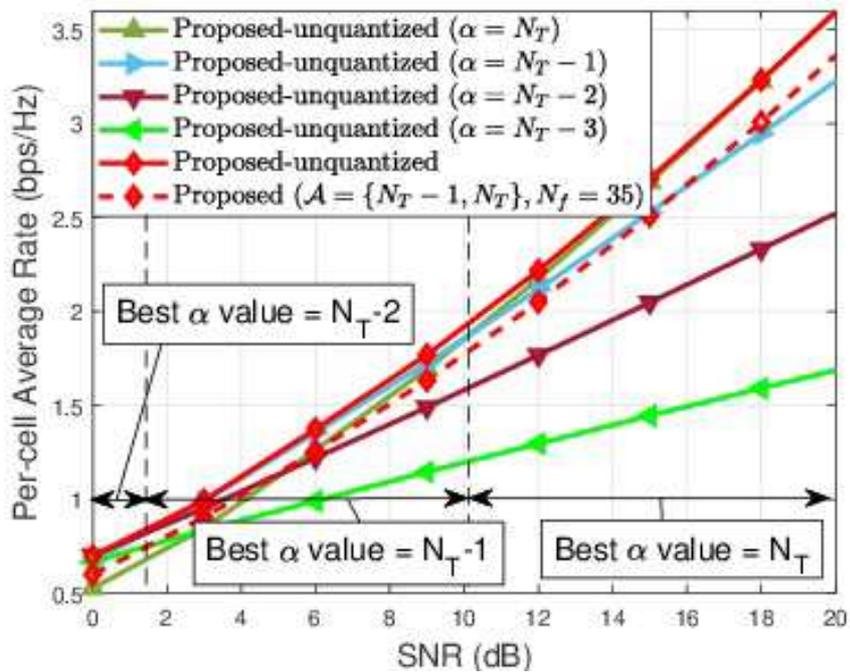}
        \caption{Per-cell average rate versus SNR of the proposed scheme compared to the unquantized versions of the proposed scheme}
        \label{fig:s1_Nt4Nc7_proposed}
    \end{subfigure}
        \caption{Per-cell average rate versus SNR for $N_T=4$ and $N_C=7$}\label{fig:s1Nt4Nc7}
\end{figure}

\begin{figure}
    \centering
    \begin{subfigure}[b]{\if\mycmd1 0.6 \else 0.21 \fi\paperwidth}
        \includegraphics[width=\textwidth]{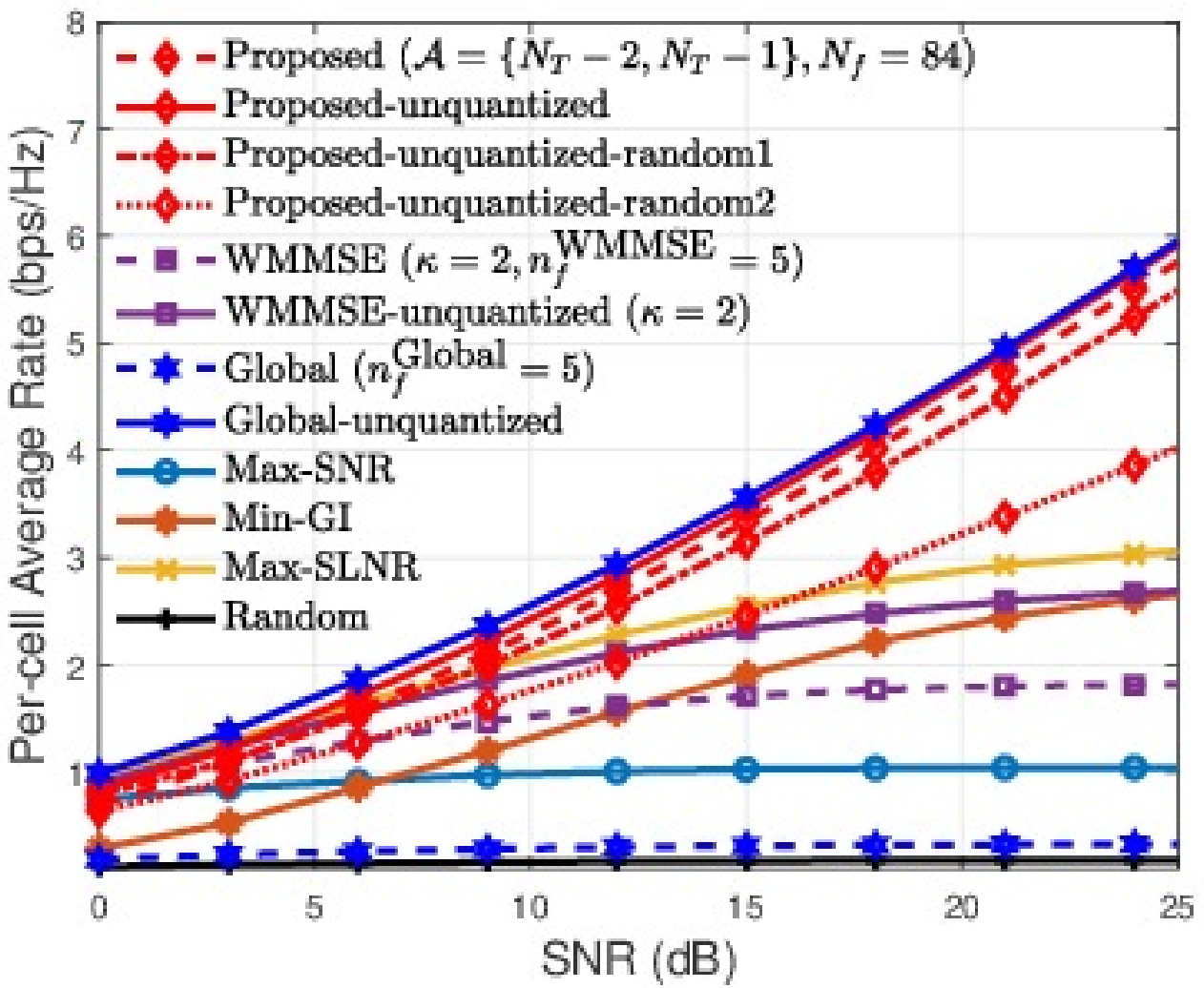}
        \caption{Per-cell average rate versus SNR of the proposed scheme compared to the existing schemes}
        \label{fig:s1_Nt8Nc9_SumRate vs. SNR}
    \end{subfigure}
        \begin{subfigure}[b]{\if\mycmd1 0.6 \else 0.21 \fi\paperwidth}
        \includegraphics[width=\textwidth]{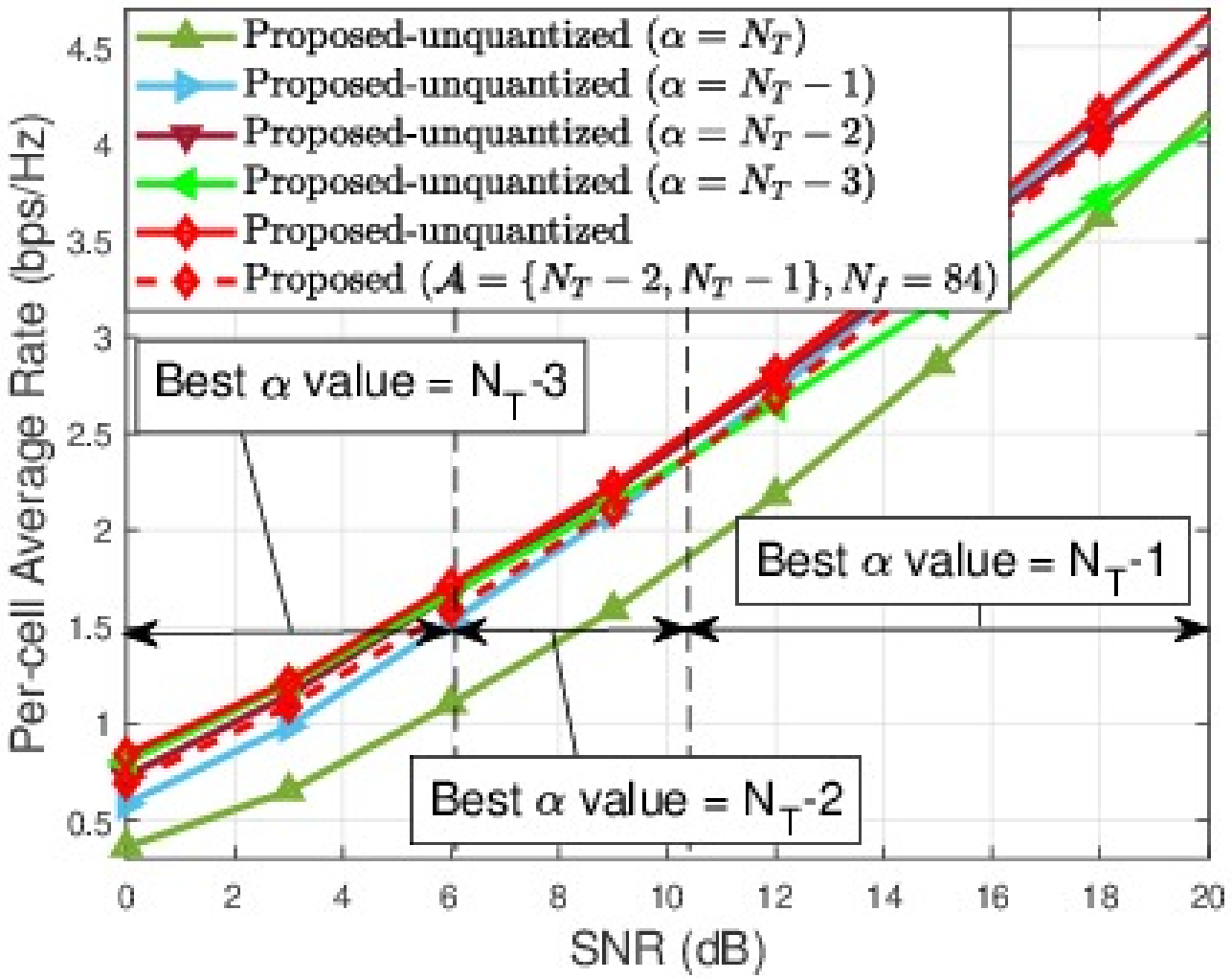}
        \caption{Per-cell average rate versus SNR of the proposed scheme compared to the unquantized versions of the proposed scheme}
        \label{fig:s1_Nt8Nc9_proposed}
    \end{subfigure}
        \caption{Per-cell average rate versus SNR for $N_T=8$ and $N_C=9$}\label{fig:s1Nt8Nc9}
\end{figure}

Figures \ref{fig:s1Nt4Nc7} and \ref{fig:s1Nt8Nc9} demonstrate the average achievable rate per-cell versus SNR for ($N_T=4$, $N_C=7$) and ($N_T=8$, $N_C=9$), respectively, under Rayleigh fading environment.
In Figs. \ref{fig:s1_Nt4Nc7_SumRate vs. SNR} and \ref{fig:s1_Nt8Nc9_SumRate vs. SNR}, the existing schemes discussed in Section \ref{sec:information exchange comparison} are compared with the proposed scheme.
For `WMMSE' and `Global,' the set of ($\kappa$, $n_f^{\textrm{WMMSE}}$, $n_f^{\textrm{Global}}$) is assumed to be (2, 2, 2) for ($N_T=4$, $N_C=7$) and (2, 5, 5) for ($N_T=8$, $N_C=9$), respectively, for fair comparison of the amount of the information exchange.
In addition, three schemes requiring only local CSI without information exchange are also considered as follows.
First, `Max-SNR' is considered, in which all the beamforming vectors are designed only to maximize the channel gain of the desired channels.
Second, `Min-GI' \cite{H_Yang17_TWC} is considered, where all the beamforming vectors are determined only to minimize GI.
Third, in `Max-SLNR' \cite{E_Bjornson10_TSP}, all the beamforming vectors are constructed maximizing SLNR.
In the baseline `Random' scheme, each beamforming vector is randomly determined.
To show the impact of the process of determination of $\alpha$ and $
\mathcal{F}$, `Proposed-unquantized-random1' and `Proposed-unquantized-random2' are considered.
In `Proposed-unquantized-random1,' $\alpha$ is selected by the proposed algorithm and $\mathcal{F}$ is randomly selected for given $\alpha$.
In `Proposed-unquantized-random2,' both $\alpha$ and $\mathcal{F}$ are randomly chosen.
For the comparison, unquantized versions of `WMMSE,' `Global,' and the proposed scheme are considered.

In Fig. \ref{fig:s1_Nt4Nc7_SumRate vs. SNR}, `Max-SLNR' shows the highest performance among the schemes which require local CSI only.
In case of `Proposed-unquantized,' it shows the per-cell average rate close to the optimal performance, i.e., `Global-unquantized,' in the SNR regime higher than 15dB.
In the figure, `Proposed ($\mathcal{A}=\{N_T-1,N_T\}$, $N_{f}=35$)' shows 6\textasciitilde{}11\% performance improvement compared to that of `Proposed-unquantized-random1'; that is, the proposed scheme has notable advantage in performance only with 35 bits of information exchange per-cell by selecting a proper set of interference-free users, $\mathcal{F}$.
On the other hand, `Proposed-unquantized-random1' shows 17\textasciitilde{}32\% per-cell average rate improvement compared with that of `Proposed-unquantized-random2,' confirming the advantage of selecting a proper number of interference-free users, $\alpha$.
In Fig. \ref{fig:s1_Nt8Nc9_SumRate vs. SNR}, the performances of `Min-GI' and `Max-SLNR' are higher than those with $N_{T}=4$, since the number of antennas is increased, resulting in lowered GI in cases of `Min-GI' and `Max-SLNR'.
With the increased number of antennas, the performance of the proposed scheme is closer to that of `Global-unquantized' than the case of $N_{T}=4$ in Fig. \ref{fig:s1_Nt4Nc7_SumRate vs. SNR}.
Because of the increased number of cells and antennas, $N_K$ is also increased; that is, the number of bits required for the information exchange is increased.
However, in Fig. \ref{fig:s1_Nt8Nc9_SumRate vs. SNR}, the proposed scheme with reasonable amount of information exchange, `Proposed ($\mathcal{A}=\{N_T-2,N_T-1\}$, $N_{f}=84$)', still shows  4\textasciitilde{}8\% improvement compared to that of `Proposed-unquantized-random1'.

In Figs. \ref{fig:s1_Nt4Nc7_proposed} and \ref{fig:s1_Nt8Nc9_proposed}, the impact of $\alpha$ is investigated by evaluating the performance of the proposed scheme but with fixed $\alpha$. As seen from the figure, the best $\alpha$ value which means the $\alpha$ value with which the proposed scheme shows the maximum per-cell average rate increases from $N_T-2$ to $N_T$ for $N_T=4$ and from $N_T-3$ to $N_T-1$ for $N_T=8$, respectively, as SNR increases; that is, the same intuition from Fig. \ref{fig:DoF_test} is confirmed with the proposed scheme.

In Fig. \ref{fig:s1_Nt4Nc7_proposedquant}, the per-cell average rates of the proposed scheme versus SNR are depicted for ($N_T=4$, $N_C=7$, $N_f=42$). For fixed $N_f$, four different sets of $(n_f, M)$ are evaluated for ($N_T=4$, $N_C=7$). In the SNR regime lower than 10dB, the proposed scheme with $\mathcal{A}=\{N_T-2,N_T-1\}$, $n_f=2$, and $M=21$ shows the maximum per-cell average rate compared to the other sets of $(n_f, M)$ for fixed $N_f$. In the SNR regime higher than 10dB, the proposed scheme with $\mathcal{A}=\{N_T-3,N_T\}$, $n_f=2$, and $M=21$ shows the maximum per-cell average rate compared with the other sets of $(n_f, M)$ for fixed $N_f$. As shown in this figure, the proper selection of $\mathcal{A}$ is crucial to maximize the per-cell average rate.

In Fig. \ref{fig:s2_Nt8Nc9101112}, the relative per-cell average rates of the proposed scheme and `Max-SLNR' normalized to the per-cell average rate of `Global-unquantized' for SNR of -5\textasciitilde{}25dB are depicted for ($N_T=8$, $N_C= 9, 10, 11, 12$).
As shown in Fig. \ref{fig:s2_Nt8Nc9101112}, `Proposed-unquantized' achieves 97\% of the per-cell average rate of `Global-unquantized', showing higher performance than the proposed schemes with fixed $\alpha$. This implies that the proposed scheme adapts $\alpha$ well for changing system parameters, e.g., $N_C$ and SNR, almost achieving the optimal performance requiring global CSI and joint beamforming vectors optimization.
It is worthwhile to note that the proposed scheme shows much higher per-cell average rate gain compared to `Max-SLNR' by finding proper weight coefficients for the SLNR equations.

Figure \ref{fig:s3_Nt8Nc9101112} shows the probability that each $\alpha$ value is chosen in the proposed scheme for ($N_T=8$, $N_C=9, 10, 11, 12$).
As shown in these figures, the proposed scheme well adapts $\alpha$ values for varying environment, showing its robustness to changes of the system parameters.

\begin{figure}
    \centering
    \begin{subfigure}[b]{\if\mycmd1 0.6 \else 0.2 \fi\paperwidth}
        \includegraphics[width=\textwidth]{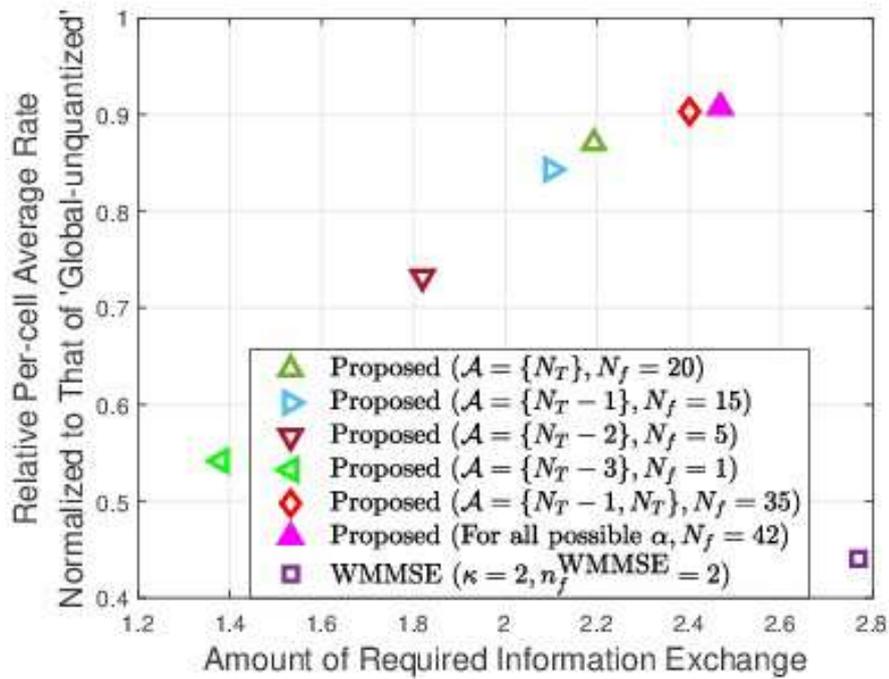}
        \caption{$N_T=4$ and $N_C=7$}
        \label{fig:s4_Nt4Nc7}
    \end{subfigure}
        \begin{subfigure}[b]{\if\mycmd1 0.6 \else 0.22 \fi\paperwidth}
        \includegraphics[width=\textwidth]{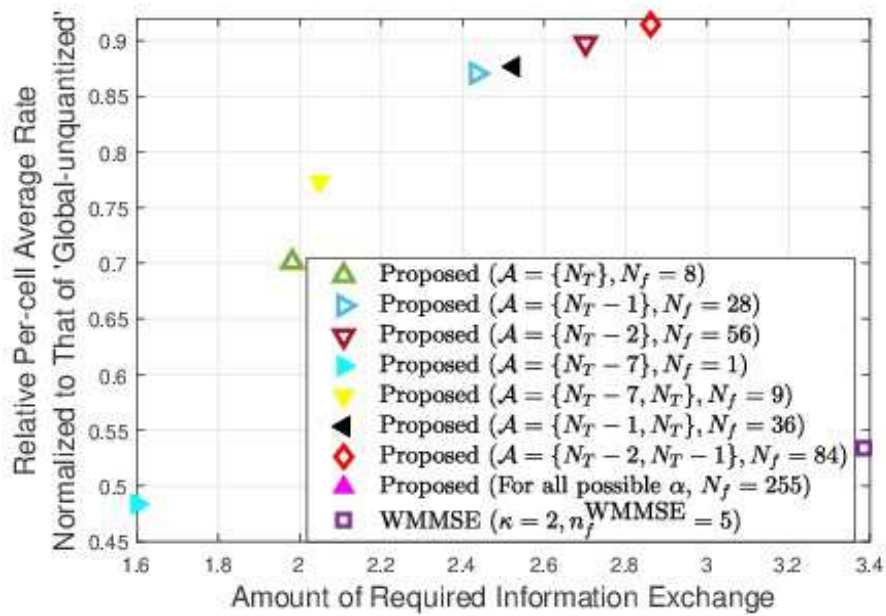}
        \caption{$N_T=8$ and $N_C=9$}
        \label{fig:s4_Nt8Nc9}
    \end{subfigure}
        \caption{Relative per-cell average rate normalized to that of `Global-unquantized' versus the amount of required information exchange for SNR of -5\textasciitilde{}25dB}\label{fig:s4}
\end{figure}

In Figs. \ref{fig:s4_Nt4Nc7} and \ref{fig:s4_Nt8Nc9}, the relative per-cell average rates normalized to the per-cell average rate of `Global-unquantized' vs. the amount of required information exchange of the proposed scheme are demonstrated compared to those of `WMMSE' in the cases of ($N_T=4$, $N_C=7$) and ($N_T=8$, $N_C=9$), respectively. Note that for a variety of $\alpha$ values the amount of information exchange and the sum-rate gain vary, obtaining a flexible trade-off between the amount of information exchange and the sum-rate.
In the case of ($N_T=4$, $N_C=7$), the per-cell average rates of the proposed schemes with $\mathcal{A}=\{N_T-1\}$ and $\mathcal{A}=\{N_T\}$ achieve 40\% and 42\% higher normalized per-cell average rate, respectively, compared to `WMMSE' even with smaller amount of information exchange.
When $\mathcal{A}=\{N_T-1,N_T\}$ is considered, the proposed scheme achieves 45\% higher normalized per-cell average rate than that of `WMMSE'.
In the case of ($N_T=8$, $N_C=9$), the proposed scheme with $\mathcal{A}=\{N_T-2,N_T-1\}$ exhibits 50\% higher rate gain than `WMMSE' with much smaller amount of information exchange.

\begin{figure}
    \centering
    \begin{subfigure}[b]{\if\mycmd1 0.6 \else 0.27 \fi\paperwidth}
        \includegraphics[width=\textwidth]{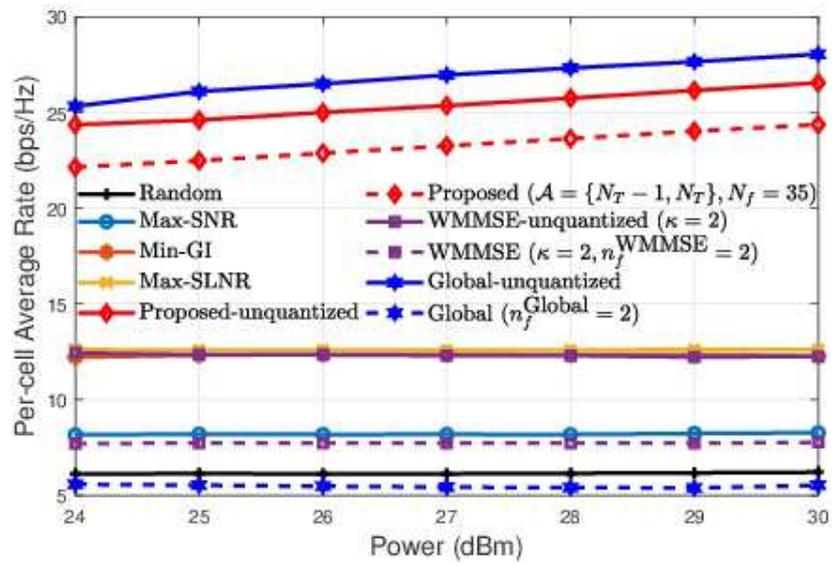}
        \caption{}
        \label{fig:s1_PL_Nt4Nc7_total}
    \end{subfigure}
        \begin{subfigure}[b]{\if\mycmd1 0.6 \else 0.16 \fi\paperwidth}
        \includegraphics[width=\textwidth]{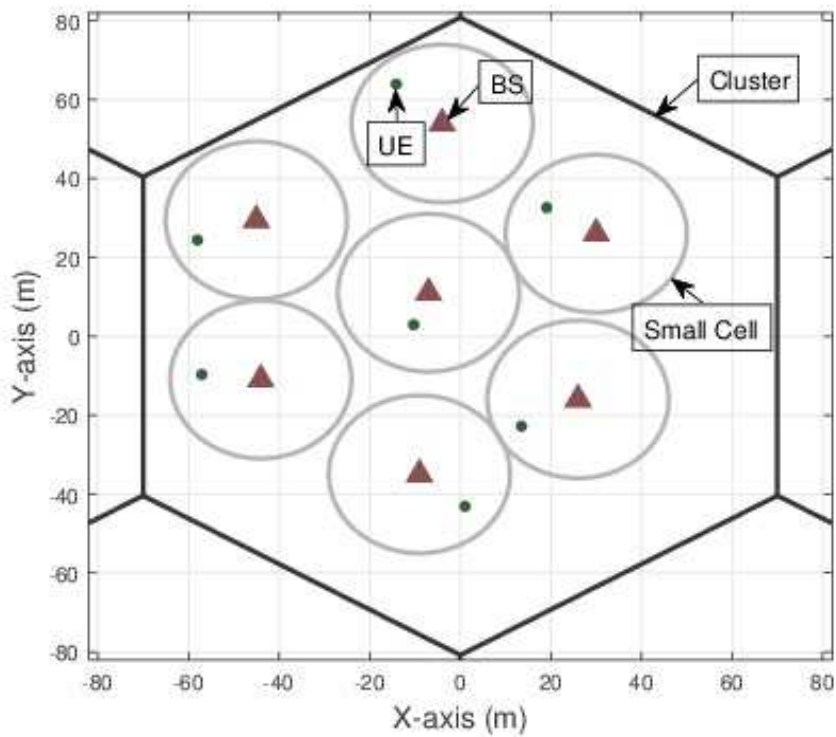}
        \caption{}
        \label{fig:PL_figure}
    \end{subfigure}
        \caption{(a) The per-cell average rate and (b) the cell configuration versus transmission power with a single user per cell for $N_T=4$ and $N_C=7$}
\label{fig:s1_PL_Nt4Nc7}\label{fig:s1_PL_Nt4Nc7}
\end{figure}

\begin{figure}
    \centering
    \begin{subfigure}[b]{\if\mycmd1 0.6 \else 0.27 \fi\paperwidth}
        \includegraphics[width=\textwidth]{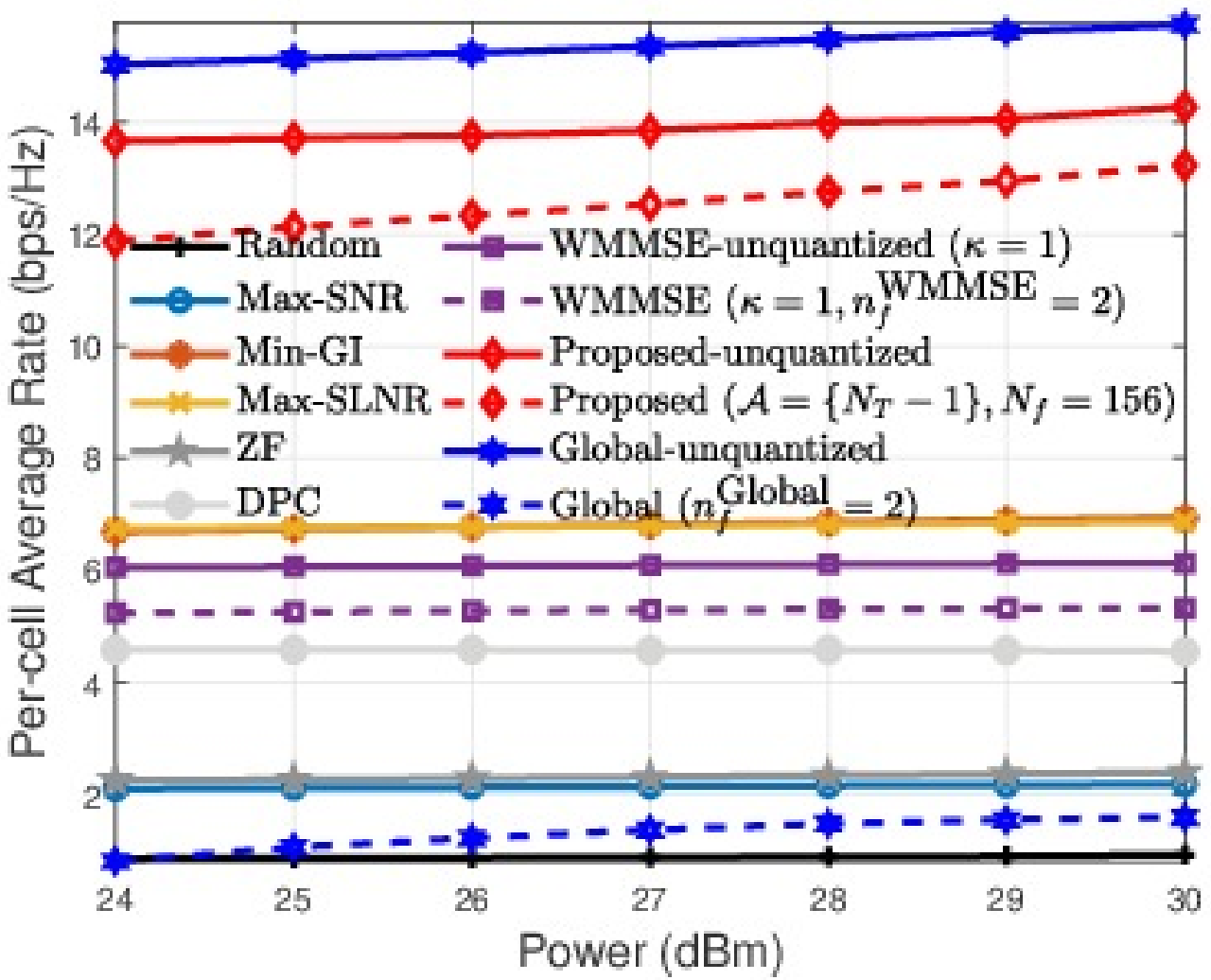}
        \caption{}
        \label{fig:s1_PL_Nt4Nc7Nu2}
    \end{subfigure}
        \begin{subfigure}[b]{\if\mycmd1 0.6 \else 0.16 \fi\paperwidth}
        \includegraphics[width=\textwidth]{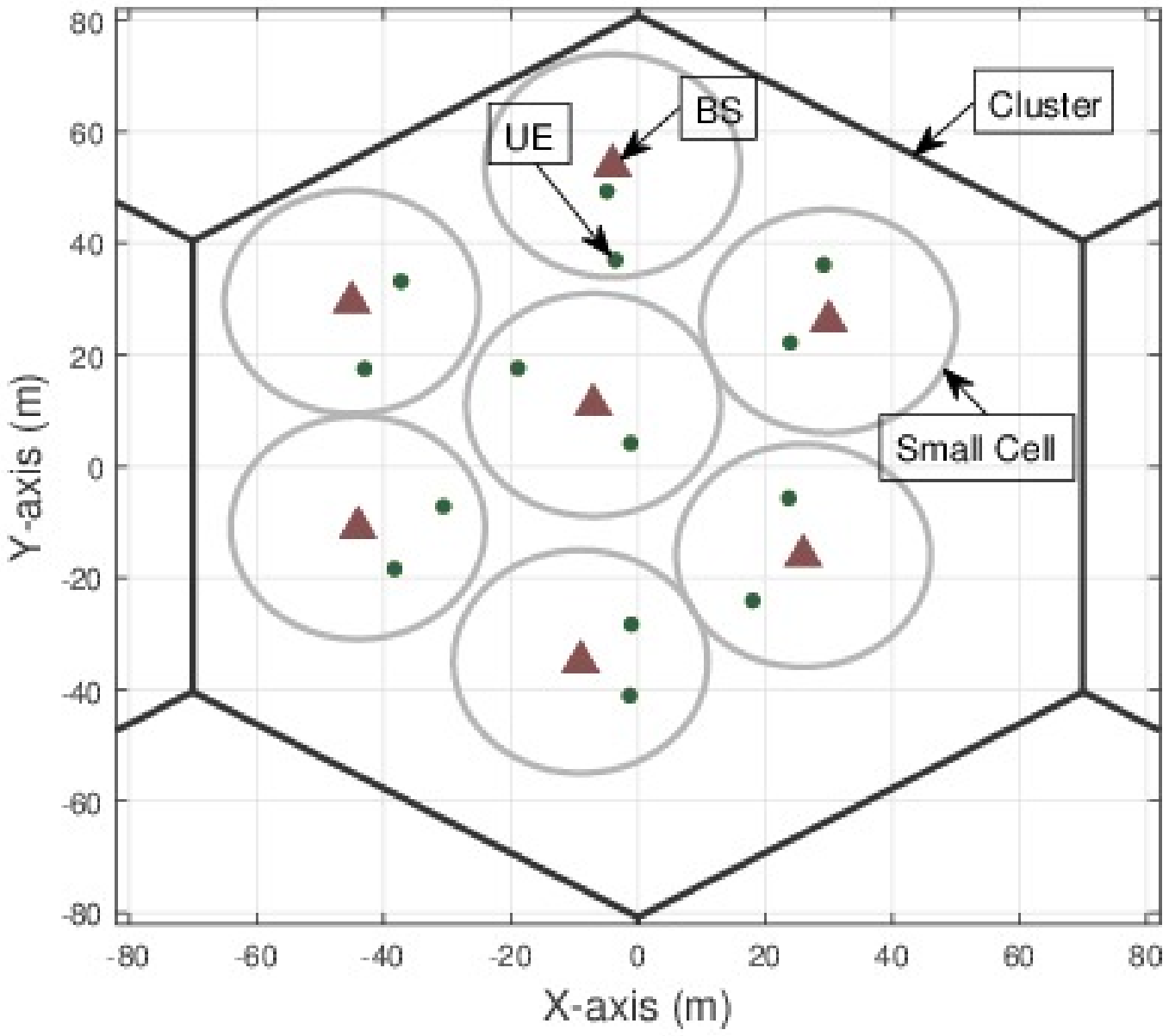}
        \caption{}
        \label{fig:PL_multiuser_figure}
    \end{subfigure}
        \caption{(a) The per-cell average rate and (b) the cell configuration versus transmission power for $N_T=4$, $N_C=7$, and $N_U=2$}
\label{fig:s1_PL_Nt4Nc7}\label{fig:s1_PL_Nt4Nc7Nu2_total}
\end{figure}

Figures \ref{fig:s1_PL_Nt4Nc7_total} and \ref{fig:s1_PL_Nt4Nc7Nu2} demonstrate the per-cell average rate versus transmission power for the single user small cell network with ($N_T=4$, $N_C=7$) and the multiuser small cell network with ($N_T=4$, $N_C=7$, $N_U=2$), respectively. Figures \ref{fig:PL_figure} and \ref{fig:PL_multiuser_figure} shows the cell configurations in small cell networks \cite{TR36.931} for \ref{fig:s1_PL_Nt4Nc7_total} and \ref{fig:s1_PL_Nt4Nc7Nu2}, respectively.
Assuming separate frequency carrier for the macro-cell BSs, e.g., Scenario 2a of the 3GPP small cell scenarios \cite{TR36.872}, there is no interference from the macro-cell BSs. Parameters and node droppings were selected from the 3GPP standards \cite{TR36.931, TR36.872} and simulation methodology therein.

As shown in Fig. \ref{fig:s1_PL_Nt4Nc7_total}, the per-cell average rates of the considered schemes except the proposed scheme and `Global-unquantized' are almost constant while that of the proposed scheme increases as the transmission power increases by mitigating intercell interference effectively. `Proposed-unquantized' and the proposed scheme with only $N_f=35$, i.e., 35 bits of information exchange per cell,  achieve about 96\% and 90\% of `Global-unquantized,' respectively, for the transmission power of 24\textasciitilde{}30dB. In Fig. \ref{fig:s1_PL_Nt4Nc7Nu2}, the zero-forcing multiuser beamforming with local CSI, labeled as `ZF', and the capacity-achieving dirty-paper coding precoding with local CSI  \cite{M_Costa83_TIT}, labeled as `DPC', are additionally evaluated for comparison. It is shown that the proposed scheme with 17 bytes, i.e., 135 bits, of information exchange per cell achieves around 94\% of `Proposed-unquantized,' while `Proposed-unquantized' achieves around 89\% of `Global-unquantized'.

\begin{center}
\begin{figure}[tbh]
\centering{}\includegraphics[width=\if\mycmd1 0.6 \else 0.22 \fi\paperwidth]{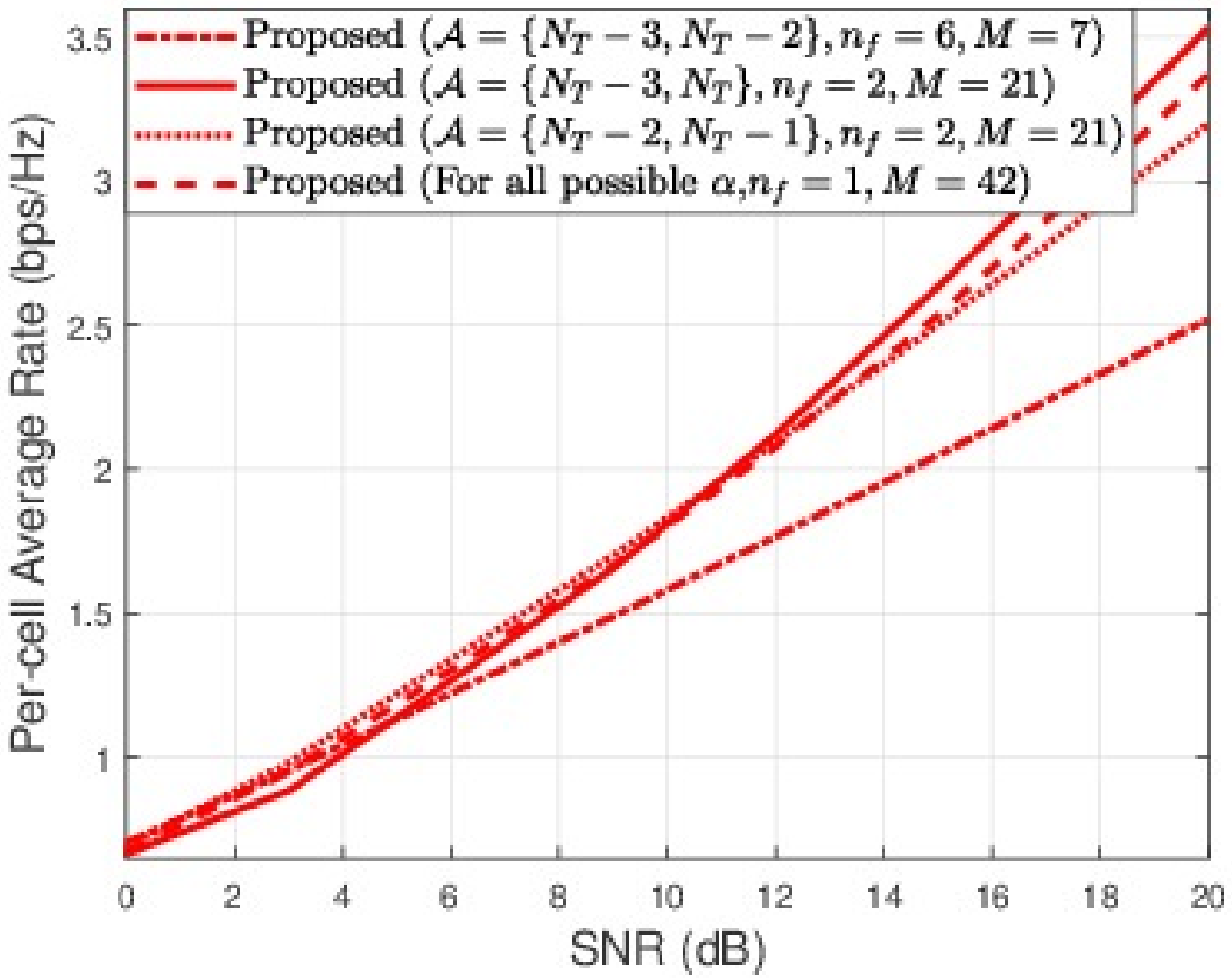}\caption{Per-cell average rate versus SNR of the proposed scheme for $N_T=4$, $N_C=7$, and $N_f=42$}\label{fig:s1_Nt4Nc7_proposedquant}
\end{figure}
\par\end{center}

\begin{center}
\begin{figure}[tbh]
\centering{}\includegraphics[width=\if\mycmd1 0.7 \else 0.35 \fi\paperwidth]{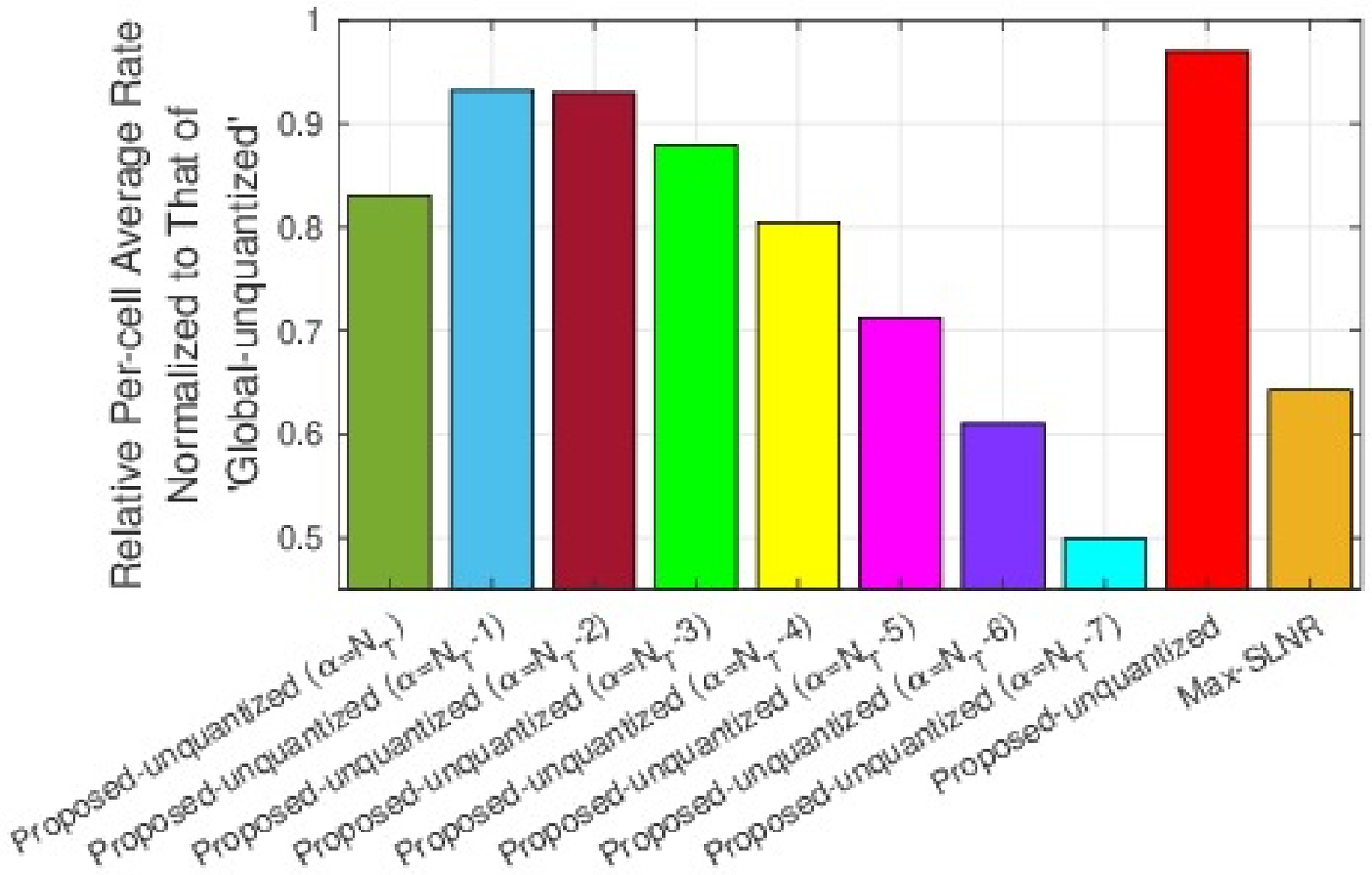}\caption{Per-cell average rate versus SNR of the proposed scheme for $N_T=8$, $N_C=9$, and $N_f=250$}\label{fig:s2_Nt8Nc9101112}
\end{figure}
\par\end{center}

\begin{center}
\begin{figure}[tbh]
\centering{}\includegraphics[width=\if\mycmd1 0.6 \else 0.25 \fi\paperwidth]{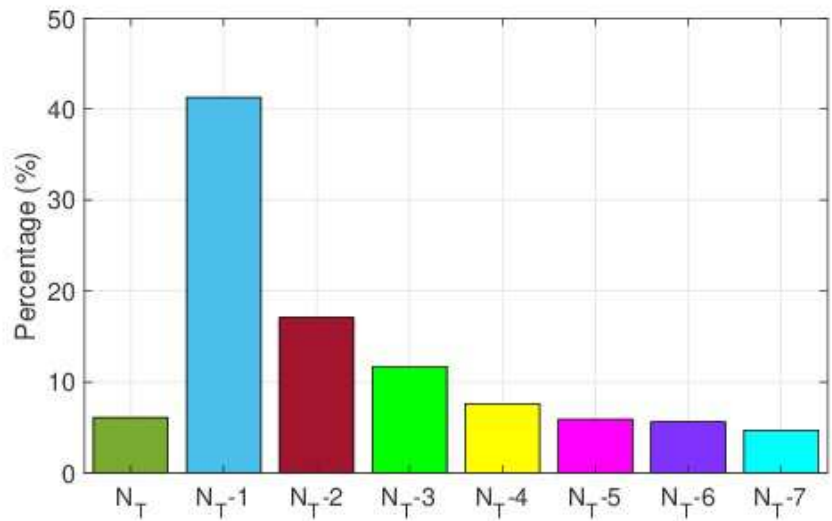}\caption{Probability that each $\alpha$ value is chosen in the proposed scheme for SNR of -5\textasciitilde{}25dB, $N_T=8$, and $N_C=9,10,11,12$}\label{fig:s3_Nt8Nc9101112}
\end{figure}
\par\end{center}

\section{Conclusion}\label{sec:conc}
We have proposed a non-iterative beamforming design scheme based on limited information exchange among the BSs to improve the sum-rate of the MISO interference channel.
Simulation results confirm that the proposed scheme closely achieves the optimal sum-rate bound, requiring less information exchange than the existing schemes.
Our future study will focus on extending the idea to the MIMO interference channel.

\bibliographystyle{IEEEtran}
\bibliography{DLBeamform2}

\end{document}